\documentclass[a4,prd,nofootinbib,preprintnumbers
,twocolumn
]{revtex4}
\usepackage{graphicx,epsf,epsfig}
\usepackage{amsmath, amssymb, amsfonts, bbm}
\usepackage{hyperref}

\usepackage{color}

\newcommand{\be}{\begin{equation}}
\newcommand{\ee}{\end{equation}}
\newcommand{\bd}{\begin{displaymath}}
\newcommand{\ed}{\end{displaymath}}
\newcommand{\bea}{\begin{eqnarray}}
\newcommand{\eea}{\end{eqnarray}}
\newcommand{\nn}{\nonumber}

\begin{document}
\title{Running soft parameters in SUSY models with multiple $U(1)$ gauge factors}
\author{Renato M. Fonseca}\email{renato.fonseca@ist.utl.pt}
\preprint{IFIC/11-34, CFTP/11-014}
\affiliation{Centro de F\'isica Te\'orica de Part\'iculas, Instituto Superior T\'ecnico,
Universidade T\'ecnica de Lisboa, Av. Rovisco Pais 1, 1049-001 Lisboa, Portugal}
\author{Michal Malinsk\'y}\email{malinsky@ific.uv.es}
\affiliation{AHEP Group, Instituto de F\'\i sica Corpuscular -- C.S.I.C./Universitat de Val\`encia Edificio de Institutos de Paterna, Apartado 22085, E--46071 Val\`encia, Spain}
\author{Werner Porod}\email{porod@physik.uni-wuerzburg.de}
\affiliation{Institut f\"ur Theoretische Physik und Astronomie, Universit\"at W\"urzburg Am Hubland, 97074 Wuerzburg}
\author{Florian Staub}\email{florian.staub@physik.uni-wuerzburg.de}
\affiliation{Institut f\"ur Theoretische Physik und Astronomie, Universit\"at W\"urzburg Am Hubland, 97074 Wuerzburg}
\begin{abstract}
We generalize the two-loop renormalization group equations for the parameters of the softly broken SUSY gauge theories given 
 in the literature
to the most general case when the gauge group contains more than
 a single abelian gauge factor. The complete method is illustrated 
at  two-loop within a specific example and compared to some of the previously proposed partial treatments.
\end{abstract}
\maketitle
\section{Introduction}
Since the advent of the renormalization group (RG) techniques
\cite{Gross:1973id,Gross:1973ju,Politzer:1973fx,Wilson:1973jj}, a lot of effort
has been put
into the calculation of the \(\beta\)-functions and anomalous dimensions of
specific theories. For instance, full-fledged two-loop formulae for
non-supersymmetric gauge models became available as early as in 1984 thanks to
the seminal works by Machacek and Vaughn 
\cite{Machacek:1983tz,Machacek:1983fi,Machacek:1984zw}. 
In the context of supersymmetry (SUSY), the need to adopt the existing
machinery for the soft SUSY-breaking sector postponed the arrival of the 
first generic two-loop results for about ten years   
\cite{Yamada:1993uh,Yamada:1993ga,Martin:1993zk,Yamada:1994id,Jack:1994kd,%
Jack:1994rk,Jack:1997eh,Jack:1998iy}. 
Since then, there have even been attempts to go beyond two loops in the literature, c.f.,  \cite{Ferreira:1996ug,Jack:2007ni}. 

For the sake of simplicity, in many of the pioneering works the gauge group was
assumed to contain at most one abelian gauge factor. The point is that with more
than a single gauged $U(1)$ in play, a new qualitative feature requiring a
dedicated treatment emerges. This is due to the fact that abelian field
tensors $F_{\mu\nu}$ are not only gauge-covariant but rather gauge-invariant
quantities and, thus, unlike the non-abelian ones, they can contract among each
other without violating gauge invariance, giving rise to off-diagonal 
 kinetic terms~\cite{Holdom:1985ag,Babu:1997st}. 

Moreover, even if such terms happen to be absent from the tree level Lagrangian
at a certain scale, they are in general re-introduced by the
renormalization-group evolution \cite{delAguila:1988jz,delAguila:1987st}. 
The reason is that the anomalous dimension $\gamma$ driving the
relevant renormalization group equations (RGEs) are in
general non-diagonal symmetric matrices in
the gauge-field space, thus giving rise to off-diagonal corrections to the gauge boson
propagators. These, in turn, require extra counterterms in order 
to retain
renormalizability. 

Actually, there are several exceptions to this basic rule. 
For instance, it can be that all the relevant $U(1)$ couplings originate from a
common gauge factor and, thus, barring threshold effects, all of them happen to
be equal at a certain scale. In such a case, accidentally, the charges and the
gauge fields can be simultaneously rotated at the one-loop level so that no
off-diagonalities pop up in~$\gamma$ \cite{delAguila:1988jz,Martin:1993zk} and
one can use the simple form of the RGEs for individual gauge couplings. 
This is relatively easy to implement in the non-SUSY case where only the gauge
sector has to be taken into account; the only price to be
paid is the presence of continuous charges in the game.

In supersymmetry, the $U(1)$ gaugino soft masses can also mix, and thus one has
to deal with the non-diagonalities in the gaugino sector too. Again, the rotated
basis can be helpful if both , gaugino masses and gauge couplings,
 unify at the same scale.  However, this method is consistent only at the one-loop level where
the evolution equations for the gauge couplings and gaugino soft masses
essentially coincide. At two loops, Yukawa couplings  and trilinear
soft SUSY breaking couplings enter and the relevant
algebraic structures are independent of
each other which, in turn, renders this approach useless.

For the non-SUSY gauge theories, the full generalization of the original two-loop results for  gauge groups with at most a single $U(1)$ factor to the case with multiple $U(1)$'s has been formulated relatively recently, see, e.g., \cite{Luo:2002iq}  and dedicated two-loop studies focusing on such effects in the context of, e.g., grand unified theories (GUTs) are available \cite{Bertolini:2009qj}.   
However, for the softly-broken SUSY gauge theories, the general two-loop
evolution equations for the  soft-breaking parameters in presence of the
$U(1)$-mixing effects have not yet been given.\footnote{ In ref.\ 
\cite{Jack:2000jr} the effect of the mixings of several $U(1)$s has
been taken into account in the anomalous dimensions of the superfields
and in the beta-functions of the gauge couplings which
serve as basis for the corresponding parts in the RGEs of the
soft SUSY breaking parameters.}

In this study, we aim to fill this gap by presenting a set of substitution rules
which generalize the results of \cite{Martin:1993zk,Yamada:1994id} to the
case where the gauge
group involves more than a single abelian gauge factor.   

The practical applications of these results are manifold. For instance, in SUSY
GUTs featuring an extended intermediate $U(1)_R \times
U(1)_{B-L}$ stage, see e.g. \cite{Malinsky:2005bi}, the $U(1)$-mixing effects
can shift the effective MSSM bino soft mass by several per cent with respect to
the na\"\i ve estimate where such effects are neglected. In principle, this 
can have non-negligible effects for the the low-energy phenomenology. 
In this respect,
let us just mention that the theories with a gauged $U(1)_{B-L}$ surviving down
to the proximity of the soft SUSY-breaking scale have become rather popular
recently
due to their interesting implications for the R-parity and the mechanism of its 
spontaneous
violation \cite{Khalil:2007dr,FileviezPerez:2010ek,Barger:2008wn}, for
Leptogenesis \cite{Pelto:2010vq,Babu:2009pi}, etc.

This work is organized as follows: In Sect.~\ref{sect-methods} we recapitulate the salient features of gauge theories with  several
 abelian gauge factor focusing namely on the different renormalization
conventions. A specific scheme in which the desired generalization of
\cite{Martin:1993zk} can be carried out in a particularly efficient way is
identified. In Sect.~\ref{sect-results} the relevant substitution rules
upgrading those in \cite{Martin:1993zk} to the most general form are given and
the methods for resolving some ambiguities emerging throughout their derivation
are briefly commented upon. In Sect.~\ref{sect:numerics} we discuss illustrate the 
the importance of the kinetic mixing effects in a pair of specific models, focusing namely on the 
comparison between the ``rotated basis'' method advocated in \cite{Martin:1993zk} and the
full-fledged two-loop treatment.
Then we conclude. 
For the sake of completeness, we add a set of appendices: some technical 
 details of the renormalization scheme definition are given in Appendix 
\ref{app:renormalizationscheme}; the basic formulae of 
\cite{Martin:1993zk}
for a simple gauge group and their  generalization to the case with product 
groups can be found in Appendix \ref{sect-rges}. In Appendix \ref{sect-appendix-results},  the interested reader can find  details of the 
derivation of our main results  presented in Sect.~\ref{sect-results}. Finally, Appendix \ref{app:matching} is devoted to several remarks on the gauge and gaugino matching in theories with multiple $U(1)$ gauge factors.

\section{Methods}\label{sect-methods}

As  mentioned in the introduction, going from a single-$U(1)$ to
the
multiple-$U(1)$ case is not straightforward as it generally amounts to a
qualitative change in the Lagrangian. In particular, there is a need for an
extra set of counterterms which, in simple words, keep the renormalized
off-diagonal two-point Green's functions in the gauge sector finite. This,
however, implies that the  renormalized Lagrangian must contain a structure
connecting the field tensors associated to different $U(1)$'s in the
gauge-kinetic terms, namely 
\be\label{eq:xi}
{\cal L}_{kin.}\ni -\tfrac{1}{4} F_{\mu\nu}\xi F^{\mu\nu}
\ee
where the different field tensors have been grouped into an $n$-dimensional
vector $F_{\mu\nu}$ (with $n$ denoting the number of independent gauged $U(1)$
factors) and $\xi$ is an $n\times n$ real and symmetric matrix. This
 amounts to
$\tfrac{1}{2}n(n-1)$ extra dynamical parameters.
These quantities are then governed by a new set of evolution equations which have to be added to those governing the individual gauge couplings and other relevant parameters such as Yukawas etc. This, indeed, is the method adopted in some of the first studies of the subject, see, e.g.,  \cite{Luo:2002iq}. 

Alternatively, one can work in a renormalization scheme in which the $\xi$-term in Eq.~(\ref{eq:xi}) is transformed out by a suitable redefinition of the gauge fields, namely, 
\be
A\to \xi^{1/2}A\,
\ee 
which also leads to the canonical normalization of the gauge
fields.
This, indeed, affects the interaction part of the covariant derivative
\be
Q_i^{T}{\tilde G} A \to Q_i^{T}{\tilde G}\xi^{-1/2}A\,,
\ee 
where $\tilde G$ is the original diagonal matrix\footnote{with indices in the
group and gauge-field spaces, respectively} of $n$ individual gauge couplings
associated to the $n$  abelian gauge factors  and  $Q_i$ is the
vector\footnote{with a lower index assigning the corresponding
matter-field} of the
relevant $U(1)$ charges. Similarly, the gauge-kinetic counterterm is transformed
\be
Z_{A}^{1/2}\xi_{B}Z_{A}^{1/2}-\xi\to
\xi^{-1/2}Z_{A}^{1/2}\xi_{B}Z_{A}^{1/2}\xi^{-1/2}-1 \; \equiv \delta
Z_{\tilde{A}},
\ee
where the subscript $B$ denotes bare  quantities and $Z_{A}^{1/2}$ is the
original (diagonal) gauge-field renormalization factor $A_{B}=Z_{A}^{1/2}A$.
Hence, the $\xi^{-1/2}$ factor can  be subsumed into a new set of  $\tfrac{1}{2}n(n-1)$ ``effective'' gauge couplings whose combinations populate the off-diagonal entries of an ``extended gauge-coupling matrix'' 
\be
G\equiv \tilde G \xi^{-1/2}\,,
\ee 
and a suitably redefined gauge-kinetic counterterm.

Thus, in this scheme, the off-diagonality in the gauge-kinetic part of the
renormalized Lagrangian is absorbed by the covariant derivative, while the
gauge-kinetic counterterm $\delta Z_{\tilde{A}}$ 
is naturally off-diagonal
in order to absorb the divergences in the off-diagonal two-point functions.
Moreover, the simple QED-like relation between the bare and renormalized abelian
gauge coupling matrices (omitting the tildes)
\be
G_{B}=G Z_{A}^{-1/2}
\ee
remains intact because the relevant Ward identities that lead to the
cancellation of $Z_{\psi}$ and the $Z_{G}$ factors, c.f., Eq.~(\ref{deltaZAtilde}),  follow from the gauge invariance. 
Therefore, it is sufficient to work with a matrix-like gauge-coupling structure forgetting entirely about the $\xi$-origin of its off-diagonal entries. 

 This strategy, which is entirely equivalent to the former one with a dynamical
$\xi$, is much more suitable for our task because it essentially amounts to
replacing all the  polynomials including individual gauge couplings in
\cite{Martin:1993zk} by the relevant matrix structures, with no
need\footnote{Obviously, no information is lost so one can obtain the relevant
RGEs for $\xi$ components from the ones with the matrix-like gauge couplings.
Indeed, the number of the off-diagonal entries in $\xi$ is the same like the
number of independent physical parameters governing the off-diagonal entries of
$G$; here one has to take into account the freedom to bring $G$ into a
triangular form by a suitable redefinition of the $U(1)$ charges.} to deal with
the evolution equations for the $\xi$ matrix not discussed here.       

This, however, is not entirely straightforward in practice. Indeed, the 
commutativity of c-numbers has been widely used in  \cite{Martin:1993zk} in 
order to cast their results in a compact form. Thus, one has to be very 
careful to avoid ambiguities stemming from the generic non-commutativity of 
the matrix-like $G$'s. Furthermore, also the abelian gaugino soft masses 
have to  be arranged into a matrix structure $M$, which brings in an 
extra complication.  

In doing so, an invaluable key is provided by some of the residual
reparametrization symmetries of the renormalized Lagrangian. In particular,
\bea
Q_i&\to&O_{1} Q_i\,,\label{O1}\\
G&\to&O_{1}GO_{2}^{T}\,,\label{O1GO2}\\
A&\to&O_{2}A\,,\label{O2}
\eea
where $O_{1}$ and $O_{2}$ are arbitrary orthogonal matrices acting in the group
and gauge-field spaces, respectively, leave the interaction part of the
covariant derivative $Q_i^{T}GA$  invariant. Under the same set of
transformations, the gaugino mass matrix is rotated to
\be
M\to O_{2}MO_{2}^{T}\,.\label{M_O2}
\ee 
Naturally, these symmetries must be reflected at the RGE level. 

Thus, for instance, only those combinations $C$ of $G$ and $\gamma \propto
\sum_i Q_i Q_i^{T}$ that transform as $C\to O_{1}CO_{2}^{T}$ are allowed
to enter the right-hand side of the renormalization group equation for $G$.
However, at one-loop level, there is only one structure involving a third power
of $G$ and one power of $\gamma$ that can come up from a matter-field  loop in
the gauge propagator, namely $GG^{T}\gamma G$, so one immediately concludes that
\be
\beta_{G}^{\rm 1 loop}\propto GG^{T}\gamma G\,.
\ee
The proportionality coefficient is trivially obtained by matching this to the single-$U(1)$ case. This also illustrates that it is more convenient to work in the scheme with off-diagonal $G$ than in the scheme with a non-trivial $\xi$, simply because the transformation properties of $G$ (which is a general real matrix) are more restrictive than the transformation properties of $\xi$ (which is symmetric). 

However, at two loop-level this becomes more complicated because then, for
instance, all gauge couplings including those corresponding to the semi-simple
part of the total gauge group mix among each other and/or with the relevant
gaugino masses. Next, different Feynman-graph topologies can be subsumed under
the same specific term in \cite{Martin:1993zk,Yamada:1994id} and, hence,
ambiguities  must be resolved, which often require some amount of a ``reverse
engineering''.

Nevertheless, as we shall demonstrate in the next section, all such
ambiguities, if properly traced back to the original diagrams, can be sorted out
and a clear and elegant picture emerges. 

\section{Results}\label{sect-results}
In this section, we shall describe the generic method of constructing the fully general two-loop RGEs for softly-broken supersymmetric gauge theories out of the results of \cite{Martin:1993zk,Yamada:1994id} relevant to the case of at most a single abelian gauge-group factor. For the sake of completeness, the relevant formulae for the cases of (i)
 a simple gauge group and (ii) the product of several simple factors with
at most a single $U(1)$ are reiterated in Appendices \ref{sect-rges-simple} and \ref{sect-rges-products}, respectively.
The computation has
been done using the $\overline{DR}'$ scheme defined in \cite{Jack:1994rk}.

\subsection{Notation and conventions}

The gauge group is taken to be $G_{A}\otimes G_{B}\otimes \ldots\otimes U\left(1\right)^{n}$,
where the $G_{X}$'s are simple groups. We shall use uppercase indices
for simple group-factors only; lowercase indices are used either for all groups or,
in some specific cases, for $U(1)$s only\footnote{This will be evident from the context; we follow as closely as possible
\cite{Martin:1993zk} and when quoting results contained therein,
the $a$ and $b$ indices go over all groups (simple and $U(1)$ groups).
On other occasions, when referring to particular components of the
$U(1)$-related $G$, $M$ and $V$ matrices and vectors, $a$ and $b$
stretch over the $U(1)$ groups only.
}. As mentioned before, the $U(1)$ sector should be treated as a whole and described in terms of a general real 
$n\times n$ gauge-coupling matrix $G$, an $n\times n$ symmetric soft-SUSY breaking gaugino mass-matrix $M$ and a column vector
of charges $Q_{i}$ for each chiral supermultiplet $\Phi_{i}$. Notice, however, that $V_{i}\equiv G^{T}Q_{i}$ for each $i$ are the only
combinations of $Q_{i}$ and $G$ which appear in the Lagrangian and, thus, all the general RGEs can be, in principle, written in terms of $V$'s and $M$ only. We shall follow this convention with a single exception of the evolution equations for the gauge couplings which are traditionally written in terms of ${\rm d}G/{\rm d}\log t$ rather than ${\rm d}V/{\rm d}\log t$ - indeed, in this case we shall adhere to the usual practice.  As a consequence, we expect an isolated $G$ popping up in these equations.

Before proceeding any further we shall define some of the expressions that are used in the RGEs:
\begin{itemize}
\item $C_{a}\left(i\right)$: Quadratic Casimir invariant of the representation
of superfield $\Phi_{i}$ under the group $G_{a}$;
\item $C\left(G_{a}\right)$: Quadratic Casimir invariant of the adjoint
representation of group $G_{a}$;
\item $S_{a}\left(i\right)$: Dynkin index of the representation of superfield
$\Phi_{i}$ under the group $G_{a}$;
\item $d_{a}\left(i\right)$: Dimension of the representation of $\Phi_{i}$
under the group $G_{a}$;
\item $d\left(G_{a}\right)$: Dimension of group $G_{a}$;
\item $S_{a}\left(R\right)$: Dynkin index of group $G_{a}$ summed over
all chiral supermultiplets - $S_{a}\left(R\right)=\sum_{i}\frac{S_{a}\left(i\right)}{d_{a}\left(i\right)}$; 
\item $S_{a}\left(R\right)C_{b}\left(R\right)$: Defined as $\sum_{i}\frac{S_{a}\left(i\right)C_{b}\left(i\right)}{d_{a}\left(i\right)}$;
\item $S_{a}\left(R\right)V_{R}^{T}V_{R}$: Defined as $\sum_{i}\frac{S_{a}\left(i\right)V_{i}^{T}V_{i}}{d_{a}\left(i\right)}$; 
\item $S_{a}\left(R\right)V_{R}^{T}MV_{R}$: Defined as $\sum_{i}\frac{S_{a}\left(i\right)V_{i}^{T}MV_{i}}{d_{a}\left(i\right)}$;
\end{itemize}
In addition, sometimes one has to deal with the explicit representation
matrices of the gauge groups (denoted in \cite{Martin:1993zk} by ${\bf t}^{Aj}_i$). Notice that here $A$ is not a group
index but rather a coordinate in the adjoint representation of the corresponding Lie algebra, (e.g., $A=1,..,3$ in $SU(2)$, $A=1,..,8$ in $SU(3)$ etc.).

Naturally, whenever we refer to  results of
refs.~\cite{Martin:1993zk,Yamada:1994id} for a simple
gauge group (collected in Appendix \ref{sect-rges-simple}), the  $a$ and $b$ indices will be omitted. In all cases, repeated indices are not implicitly summed over.

\subsection{Constructing the general substitution rules}

Let us now sketch in more detail the general strategy for upgrading the ``product'' substitution rules of Sect.~III in ref.\
 \cite{Martin:1993zk} to the most general case of an arbitrary gauge group.
For sake of simplicity, we shall focus on a limited number of terms here; 
the interested reader can find a more elaborate exemplification of the basic procedure in Appendix \ref{sect-appendix-results}.
 
Let us begin with, e.g., the term $g^{2}C\left(r\right)$ appearing for instance in Eq.~\ref{example1} and, subsequently, in the substitution 
rules of \cite{Martin:1993zk}
for product groups, Eq.~\ref{example1product}. It is clear that this has to be replaced
by $\sum_{A}g_{A}^{2}C_{A}\left(r\right)+\text{`}U(1) \text{ part'}$. For
a single $U(1)$, $g^{2}C\left(r\right)=g^{2}y_{r}^{2}\sim V_{r}V_{r}$
so this `$U(1)$ part' can only take the form\footnote{If there is a single abelian factor group, we denote by $y_i$ the (hyper)charge of chiral superfield $\Phi_{i}$, which is just a number}:
$V_{r}^{T}V_{r}=Q_{r}^{T}GG^{T}Q_{r}$.
There is no other way to obtain a number from two vectors $V_{r}$.
Remarkably, this expression sums automatically the contributions
of all the $U(1)$'s.

Similarly, $Mg^{2}C\left(r\right)$ (in Eq.~\ref{example4} for example) is replaced by $\sum_{A}M_{A}g_{A}^{2}C_{A}\left(r\right)+\text{`}U(1) \text{ part'}$; the ingredients for the construction of the `$U(1)$ part' are two vectors $V_{r}$ and
the gaugino mass matrix $M$. Only $V_{r}^{T}MV_{r}$ forms a number.

In fact, this simple procedure allows us to generalize many of the
terms in the RGEs of \cite{Martin:1993zk,Yamada:1994id}, Sect.~II (and/or Appendix \ref{sect-rges-simple}). 
As a more involved example, consider for instance the $g^{4}{\bf t}_{i}^{Aj}\textrm{Tr}\left[{\bf t}^{A}C\left(r\right)m^{2}\right]$
structure popping up in Eq.~(B.25). It is not difficult to see that
all terms where the representation matrices ${\bf t}^{A}$ appear
explicitly are zero unless $A$ corresponds to an abelian group. Hence, if for
a single $U(1)$ one has $g^{4}{\bf t}_{i}^{Aj}\textrm{Tr}\left[{\bf t}^{A}C\left(r\right)m^{2}\right]=g^{4}\delta_{i}^{j}y_{i}\sum_{p}y_{p}\left[\sum_{B}g_{B}^{2}C_{B}\left(p\right)+y_{p}^{2}\right]\left(m^{2}\right)_{p}^{p}$,
it can be immediately deduced that, in the general case, $g^{4}{\bf t}_{i}^{Aj}\textrm{Tr}\left[{\bf t}^{A}C\left(r\right)m^{2}\right]\rightarrow\delta_{i}^{j}\sum_{p}\left(V_{i}^{T}V_{p}\right)\left[\sum_{B}g_{B}^{2}C_{B}\left(p\right)+\left(V_{p}^{T}V_{p}\right)\right]\left(m^{2}\right)_{p}^{p}$.
The RGEs of $G$ and $M$ represent a bigger challenge, because they
are matrix equations (i.e., the gauge indices remain open). On the other hand, this should be viewed as an advantage because all the relevant equations must then respect the reparametrization  
symmetries (\ref{O1})-(\ref{M_O2}). In this respect, let us reiterate Eqs.~(\ref{O1})-(\ref{O2}) which imply that $V$'s transform as 
$
V_{i}\rightarrow O_{2}V_{i}
$.
These symmetries are especially powerful in the $\beta$-functions for the gauge couplings which, due to Eq.~(\ref{O1GO2}), inevitably   
take the generic form
$GV_{i}\left(\cdots\right)V_{j}^{T}$ for some chiral indices $i,j$.
For example, $g^{3}S\left(R\right)\sim g^{3}\sum_{p}y_{p}^{2}$ can
only take the form $G\sum_{p}V_{p}V_{p}^{T}$.

Concerning the gaugino soft masses $M$, let us for instance take a look at the $2 g^2 S\left(R\right)M$ term
appearing in Eq.~(\ref{example5}). Its generalized variant should be, obviously, built out of a pair of $V_{p}$ vectors and the $M$ matrix. However, there are only two combinations of these objects that transform correctly under $O_{2}$, namely, $MV_{p}V_{p}^{T}$ and $V_{p}V_{p}^{T}M$. Thus, due to the symmetry of $M$, one obtains $2 g^2 S\left(R\right)M\rightarrow
M\sum_{p}V_{p}V_{p}^{T}+\sum_{p}V_{p}V_{p}^{T}M$.

Another important ingredient of the analysis is provided by the existing substitution rules linking the case of a simple gauge group (Sect.~II in \cite{Martin:1993zk} and/or Appendix~\ref{sect-rges-simple}) to the settings with group products (Sect.~III in \cite{Martin:1993zk} and/or Appendix~\ref{sect-rges-products}). Consider, for example, the $g^{5}S\left(R\right)C\left(R\right)$ term in (Eq.~\ref{g5SRCR}) which,
according to \cite{Martin:1993zk}, gets replaced by $\sum_{b}g_{a}^{3}g_{b}^{2}S_{a}\left(R\right)C_{b}\left(R\right)$, see formula (\ref{g5SRCRsubst})
for the product groups. 
Let us recall that the expression
$S\left(R\right)C\left(R\right)$ has a very particular meaning -
it is the sum of the Dynkin indices weighted by the quadratic Casimir
invariant, so $\sum_{b}g_{a}^{3}g_{b}^{2}S_{a}\left(R\right)C_{b}\left(R\right)=\sum_{b,p}g_{a}^{3}g_{b}^{2}\frac{S_{a}\left(p\right)C_{b}\left(p\right)}{d_{a}\left(p\right)}$.
With this in mind, whenever $a$ refers to the abelian part of the gauge
group, one should replace $g_{a}^{3}S_{a}\left(p\right)\rightarrow GV_{p}V_{p}^{T}$,
$\sum_{b}g_{b}^{2}C_{b}\left(p\right)\rightarrow\sum_{B}g_{B}^{2}C_{B}\left(p\right)+V_{p}^{T}V_{p}$
and $d_{a}\left(p\right)=1$. Therefore, for the abelian sector,
$g^{5}S\left(R\right)C\left(R\right)\rightarrow\sum_{p}GV_{p}V_{p}^{T}\left[
\sum_{B}g_{B}^{2}C_{B}\left(p\right)+V_{p}^{T}V_{p}\right]$.

However, sometimes even a detailed inspection of the underlying expressions does not admit for an unambiguous identification of its generalized form. Then, a careful analysis of the structure of the contributing Feynman diagrams is necessary. Remarkably, the number of such singular cases is rather limited and can be carried out rather efficiently, as
shown in  Appendix \ref{sect-appendix-results}.

\subsection{List of substitution rules}
Depending on the group sector (abelian or simple), we get different RGEs for the gauge couplings and the gaugino masses. The parameters are then either the matrices $G$, $M$ or the numbers $g_A$, $M_A$. For the abelian sector, one obtains:
\begin{align}
C\left(G\right) & \rightarrow0\,,\label{rule31}\\
g^{3}S\left(R\right) & \rightarrow G\sum_{p}V_{p}V_{p}^{T}\label{rule32}\\
g^{5}S\left(R\right)C\left(R\right) & \rightarrow\sum_{p}GV_{p}V_{p}^{T}\Bigl[\sum_{B}g_{B}^{2}C_{B}\left(p\right)+V_{p}^{T}V_{p}\Bigr]\label{rule33}\\
\frac{g^{3}C\left(k\right)}{d\left(G\right)} & \rightarrow GV_{k}V_{k}^{T}\label{rule35}\\
2g^{2}S\left(R\right)M & \rightarrow M\sum_{p}V_{p}V_{p}^{T}+\sum_{p}V_{p}V_{p}^{T}M\label{rule36}\\
g^{2}C\left(k\right) & \rightarrow V_{k}V_{k}^{T}\label{rule39}\\
2g^{2}C\left(k\right)M & \rightarrow MV_{k}V_{k}^{T}+V_{k}V_{k}^{T}M\label{rule40}\end{align}
\begin{widetext}
\begin{align}
16g^{4}S\left(R\right)C\left(R\right)M & \rightarrow\sum_{p}\left\{ 4\left(MV_{p}V_{p}^{T}+V_{p}V_{p}^{T}M\right)\left[\sum_{B}g_{B}^{2}C_{B}\left(p\right)+V_{p}^{T}V_{p}\right]\right.\nonumber \\
 & \left.+8V_{p}V_{p}^{T}\left[\sum_{B}M_{B}g_{B}^{2}C_{B}\left(p\right)+V_{p}^{T}MV_{p}\right]\right\} \label{rule37}
\end{align}

For a simple  group factor $G_{A}$, the substitution rules of \cite{Martin:1993zk}
do not need to be changed except for two cases:\begin{align}
g^{5}S\left(R\right)C\left(R\right) & \rightarrow g_{A}^{3}S_{A}\left(R\right)\left[\sum_{B}g_{B}^{2}C_{B}\left(R\right)+V_{R}^{T}V_{R}\right]\label{rule34}\\
16g^{4}S\left(R\right)C\left(R\right)M & \rightarrow8 g_A^2 M_{A}S_{A}\left(R\right)\left[\sum_{B}g_{B}^{2}C_{B}\left(R\right)+V_{R}^{T}V_{R}\right]+8 g_A^2 S_{A}\left(R\right)\left[\sum_{B}M_{B}g_{B}^{2}C_{B}\left(R\right)+V_{R}^{T}MV_{R}\right]\label{rule38}\end{align}

As for the rest of the parameters in a SUSY model, the relevant substitution
rules read:\begin{align}
g^{2}C\left(r\right) & \rightarrow\sum_{A}g_{A}^{2}C_{A}\left(r\right)+V_{r}^{T}V_{r}\label{rule14}\\
Mg^{2}C\left(r\right) & \rightarrow\sum_{A}M_{A}g_{A}^{2}C_{A}\left(r\right)+V_{r}^{T}MV_{r}\label{rule15}\\
M^{*}g^{2}C\left(r\right) & \rightarrow\sum_{A}M_{A}^{*}g_{A}^{2}C_{A}\left(r\right)+V_{r}^{T}M^{\dagger}V_{r}\label{rule16}\\
MM^{*}g^{2}C\left(r\right) & \rightarrow\sum_{A}M_{A}M_{A}^{*}g_{A}^{2}C_{A}\left(r\right)+V_{r}^{T}MM^{\dagger}V_{r}\label{rule17}\\
g^{4}C\left(r\right)S\left(R\right) & \rightarrow\sum_{A}g_{A}^{4}C_{A}\left(r\right)S_{A}\left(R\right)+\sum_{p}\left(V_{r}^{T}V_{p}\right)^{2}\label{rule18}\\
Mg^{4}C\left(r\right)S\left(R\right) & \rightarrow\sum_{A}M_{A}g_{A}^{4}C_{A}\left(r\right)S_{A}\left(R\right)+\sum_{p}\left(V_{r}^{T}MV_{p}\right)\left(V_{r}^{T}V_{p}\right)\label{rule19}\\
g^{4}C^{2}\left(r\right) & \rightarrow\sum_{A,B}g_{A}^{2}g_{B}^{2}C_{A}\left(r\right)C_{B}\left(r\right)+2\sum_{A}g_{A}^{2}C_{A}\left(r\right)\left(V_{r}^{T}V_{r}\right)+\left(V_{r}^{T}V_{r}\right)^{2}\label{rule20}\\
Mg^{4}C^{2}\left(r\right) & \rightarrow\sum_{A,B}M_{A}g_{A}^{2}g_{B}^{2}C_{A}\left(r\right)C_{B}\left(r\right)\!+\!\sum_{A}g_{A}^{2}C_{A}(r)[M_{A}(V_{r}^{T}V_{r})+(V_{r}^{T}MV_{r})]+(V_{r}^{T}MV_{r})(V_{r}^{T}V_{r})\label{rule21}\\
g^{4}C\left(G\right)C\left(r\right) & \rightarrow\sum_{A}g_{A}^{4}C\left(G_{A}\right)C_{A}\left(r\right)\label{rule22}\\
Mg^{4}C\left(G\right)C\left(r\right) & \rightarrow\sum_{A}M_{A}g_{A}^{4}C\left(G_{A}\right)C_{A}\left(r\right)\label{rule23}\\
MM^{*}g^{4}C\left(G\right)C\left(r\right) & \rightarrow\sum_{A}M_{A}M_{A}^{*}g_{A}^{4}C\left(G_{A}\right)C_{A}\left(r\right)\label{rule24}\\
g^{2}{\bf t}^{Aj}_i\textrm{Tr}\left({\bf t}^{A}m^{2}\right) & \rightarrow\delta_{i}^{j}\sum_{p}\left(V_{i}^{T}V_{p}\right)\left(m^{2}\right)_{p}^{p}\label{rule25}\\
g^{2}{\bf t}^{Aj}_i\left({\bf t}^{A}m^{2}\right)_{r}^{l} & \rightarrow\delta_{i}^{j}\left(V_{l}^{T}V_{i}\right)\left(m^{2}\right)_{r}^{l}\label{rule26}\\
g^{4}{\bf t}^{Aj}_i\textrm{Tr}\left[{\bf t}^{A}C\left(r\right)m^{2}\right] & \rightarrow\delta_{i}^{j}\sum_{p}\left(V_{i}^{T}V_{p}\right)\left[\sum_{B}g_{B}^{2}C_{B}\left(p\right)+\left(V_{p}^{T}V_{p}\right)\right]\left(m^{2}\right)_{p}^{p}\label{rule27}\\
g^{4}C\left(i\right)\textrm{Tr}\left[S\left(r\right)m^{2}\right] &
\rightarrow\sum_{A}g_{A}^{4}C_{A}\left(i\right)\textrm{Tr}\left[S_{A}\left(r\right)m^{2}\right]+\sum_{p}\left(V_{i}^{T}V_{p}\right)^{2}\left(m^{2}\right)_{p}^{p}\label{rule28}
\end{align}
\begin{align}
24g^{4}MM^{*}C\left(i\right)S\left(R\right) & \rightarrow24\sum_{A}g_{A}^{4}M_{A}M_{A}^{*}C_{A}\left(i\right)S_{A}\left(R\right)+8\sum_{p}\left[\left(V_{i}^{T}MV_{p}\right)\left(V_{i}^{T}M^{\dagger}V_{p}\right)\right.\nonumber \\
 & \quad \left.+\left(V_{i}^{T}MM^{\dagger}V_{p}\right)\left(V_{i}^{T}V_{p}\right)+\left(V_{i}^{T}M^{\dagger}MV_{p}\right)\left(V_{i}^{T}V_{p}\right)\right]\label{rule29}\\
48g^{4}MM^{*}C\left(r\right)^{2} & \rightarrow\sum_{A,B}g_{A}^{2}g_{B}^{2}C_{A}\left(r\right)C_{B}\left(r\right)\left[32M_{A}M_{A}^{*}+8M_{A}M_{B}^{*}+8M_{B}M_{A}^{*}\right]\nonumber \\
 & \quad +\sum_{A}g_{A}^{2}C_{A}\left(r\right)\left[32M_{A}M_{A}^{*}\left(V_{r}^{T}V_{r}\right)+16M_{A}\left(V_{r}^{T}M^{\dagger}V_{r}\right)+16M_{A}^{*}\left(V_{r}^{T}MV_{r}\right)+32\left(V_{r}^{T}MM^{\dagger}V_{r}\right)\right]\nonumber \\
 & \quad +\left[32\left(V_{r}^{T}MM^{\dagger}V_{r}\right)\left(V_{r}^{T}V_{r}\right)+16\left(V_{r}^{T}MV_{r}\right)\left(V_{r}^{T}M^{\dagger}V_{r}\right)\right]\label{rule30}\end{align}
\end{widetext}
\section{Comparison with other methods to include $U(1)$-mixing}
\label{sect:numerics}
\subsection{General discussion\label{sect:general}}
So far, several approaches to the SUSY $U(1)$-mixing conundrum have been proposed in the literature. Let us take a brief look at some of them and comment on their limitations as compared to the complete two-loop treatment advocated in this work.

As we have already mentioned in the Introduction, one can attempt to choose a convenient pair of bases in the $U(1)$-charge and gauge-field spaces for which the situation might simplify \cite{delAguila:1988jz,Martin:1993zk}. For instance, it is always possible to diagonalize the one-loop anomalous dimensions 
\begin{equation}
\gamma =  \sum_i Q_i Q_i^T
\end{equation}
by means of a suitable $O_{1}$ rotation $Q_i\to O_{1}Q_i\equiv Q_i'$, see
(\ref{O1}), so that $\gamma'=O_{1}\gamma O_{1}^{T}$ is diagonal. This, of
course, inflicts a change on the gauge-coupling matrix $G\to O_{1}G$. However,
if all the relevant $U(1)$ gauge couplings happen to emanate from a single
point, i.e., $G\propto 1$ at some scale, $O_{1}$ can be passed through $G$ and
absorbed by a suitable redefinition of the gauge fields (\ref{O2}) where now
$O_{2}=O_{1}$. This way, the one-loop evolution of $G$ is driven by a diagonal
$\gamma'$ and the initial condition $G\propto 1$ remains intact. Thus, no
off-diagonalities emerge in this case and it is consistent to work with the
usual RGEs for individual gauge couplings, one per each $U(1)$ factor.    

This approach, however, is generally limited to the evolution with a complete
$U(1)$ unification. This is very often not the case in practice, in particular
in the GUTs in which the hypercharge is a non-trivial linear combination of the
relevant Cartans, such as in  left-right models based on the
$SU(2)_{L}\otimes SU(2)_{R}\otimes U(1)_{B-L}$ gauge group, see Sect.~\ref{sect:numericsMRV}. Moreover, not only
gauge couplings but also the $U(1)$ gaugino soft masses should  coincide at the
unification scale otherwise the method fails in the soft sector already at the
one-loop level. The point is that only then the generalized one-loop correlation
between the gauge couplings and the gaugino masses 
\begin{equation}\label{generalizedgauginogaugecorrelation}
 G M^{-1} G^T = \text{const.}
\end{equation}
ensures the gaugino mass diagonality along the unification trajectory.  
 
At the two-loop level more complicated structures such 
as higher powers of charges,
gauge couplings, Yukawas, etc., enter the anomalous dimensions and, in general,
there is no way to diagonalize simultaneously all the evolution equations.
Though there is still a trick one can implement in the gauge sector if the
$U(1)$ couplings do not unify \cite{delAguila:1988jz}, there is no general way
out in the supersymmetric case for the gauginos as also discussed in
\cite{Braam:2011xh}. Thus, a full-fledged two-loop
approach as presented in this work is mandatory and, in fact, it turns out 
to be even technically indispensable if there happen to be more than two
abelian gauge groups as, for instance, in
\cite{delAguila:1987st}, \cite{Perez:2011dg} and many string-inspired constructions. 

\subsection{Simple illustrations\label{sect:numericsMRV}}
Let us illustrate the importance of the kinetic mixing effects in a couple of simple 
scenarios which exhibits all the salient features discussed above. 
\subsubsection{One-loop effects}
\paragraph{Gauge couplings:} 
We shall 
consider the one-loop evolution of the gauge couplings in the SUSY $SO(10)$ 
model of ref.\ \cite{Malinsky:2005bi} in which the unified gauge 
symmetry is broken down to the MSSM in three steps, namely, $SO(10)\to 
SU(3)_{c}\otimes SU(2)_{L}\otimes SU(2)_{R}\otimes U(1)_{B-L}\to 
SU(3)_{c}\otimes SU(2)_{L}\otimes U(1)_{R}\otimes U(1)_{B-L}\to {\rm 
MSSM}$; the corresponding breaking scales shall be denoted by $M_{\rm G}$, 
$M_{\rm R}$ and $M_{\rm BL}$, respectively. Further details including 
the field contents at each of the symmetry breaking stages can be found in 
ref.\ \cite{Malinsky:2005bi}. 

For our purposes, it is crucial that in this model the
ratio $M_{\rm R} /M_{\rm BL}$ can be as large as $10^{10}$ and, hence, 
the $U(1)$ mixing effects become  important.
Note that even a short 
$SU(3)_{c}\otimes SU(2)_{L}\otimes SU(2)_{R}\otimes U(1)_{B-L}$ stage is 
sufficient to split the $g_{R}$ and the $g_{B-L}$ gauge 
couplings such  that the extended gauge-coupling matrix $G$ at the 
$M_{\rm R}$ scale is rather far from being proportional to  the 
unit matrix. Thus, there is no way to choose the $O_{1}$ and $O_{2}$ 
rotation  matrices such that both $G$ and $\gamma$
\be
\gamma=N\left(\begin{array}{cc}
15/2 & -1\\
-1 & 18
\end{array}\right)N 
\ee
are simultaneously diagonalized. Here 
$N={\rm diag}(1,\sqrt{3/8})$ ensures the canonical normalization of the 
$B-L$ charge within the $SO(10)$ framework.
Therefore, the one-loop evolution equation relevant to the 
$U(1)_{R}\otimes U(1)_{B-L}$ stage has to be  matrix-like and reads
 in the abelian sector
\be
\frac{\rm d}{{\rm d}t}A^{-1}=-\gamma\,,
\ee
where $A^{-1}=4\pi (GG^{T})^{-1}$ and $t=\log(\mu/\mu_{0})/2\pi$. 
 
The reason that the $U(1)_{R}\otimes U(1)_{B-L}$ stage can be so long has to do with the fact that this gauge symmetry is broken by neutral components of an $SU(2)_{R}$ doublet pair, namely, $(1,1,+\tfrac{1}{2},-1)\oplus (1,1,-\tfrac{1}{2},+1)=\chi_{R}^{0}\oplus \overline{\chi}_{R}^{0}$ which are full SM singlets, and as such they do not affect the low-energy value of $\alpha_{Y}^{-1}$. Indeed, the would-be change inflicted on $\alpha_{Y}^{-1}$ by the presence or absence of $\chi_{R}^{0}\oplus \overline{\chi}_{R}^{0}$ is given by
\be\label{Deltaalphainverse}
\Delta\alpha^{-1}_{Y}=p_{Y}^{T}\,\Delta A^{-1}(M_{\rm BL})\, p_{Y}\propto p_{Y}^{T}\,\Delta\gamma\, p_{Y}=0
\ee
where $p_{Y}^{T}=(\sqrt{3/5},\sqrt{2/5})$ are the coordinates of the MSSM hypercharge in the $U(1)_{R}\otimes U(1)_{B-L}$ algebra and $\Delta\gamma$ denotes the relevant change of the  $\gamma$ matrix.
Therefore, at the one-loop level, the position of the $M_{\rm BL}$ scale is not constrained by the low-energy data and, hence, barring other phenomenological constraints, it can be pushed as close to the MSSM scale $M_{S}$ as desired.
\begin{figure}[t]
\includegraphics[width=8.5cm]{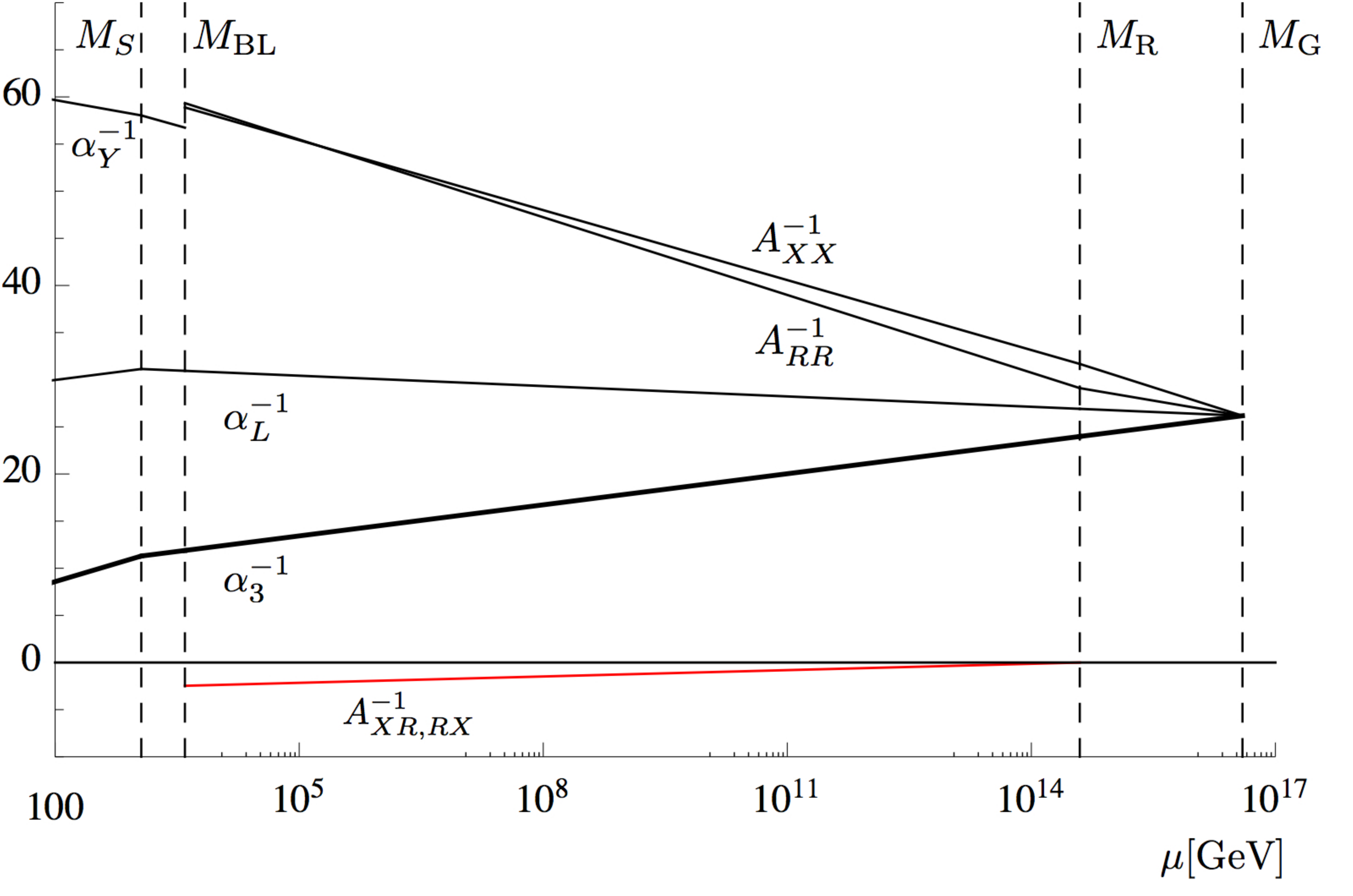}
\caption{One-loop gauge-coupling evolution in the MRV model \cite{Malinsky:2005bi}. The position of the GUT scale, the unified gauge coupling and the intermediate symmetry-breaking scale $M_{R}$ were chosen in such a way to fit the electroweak data with $\alpha^{-1}_{Y}(M_{Z})=59.73$. The close-to-zero red line in the $M_{BL}-M_{R}$ domain depicts the evolution of the off-diagonal entries of the  $A^{-1}=4\pi (GG^{T})^{-1}$ matrix which, at the one-loop level, scales linearly with $\log\mu$. The ``optical discontinuity'' in $\alpha_{Y}^{-1}$ at the $M_{\rm BL}$ scale  owns to the generalized matching condition (\ref{matching}). \label{MRVrunningOK}}
\end{figure}
\begin{figure}[t]
\includegraphics[width=8.5cm]{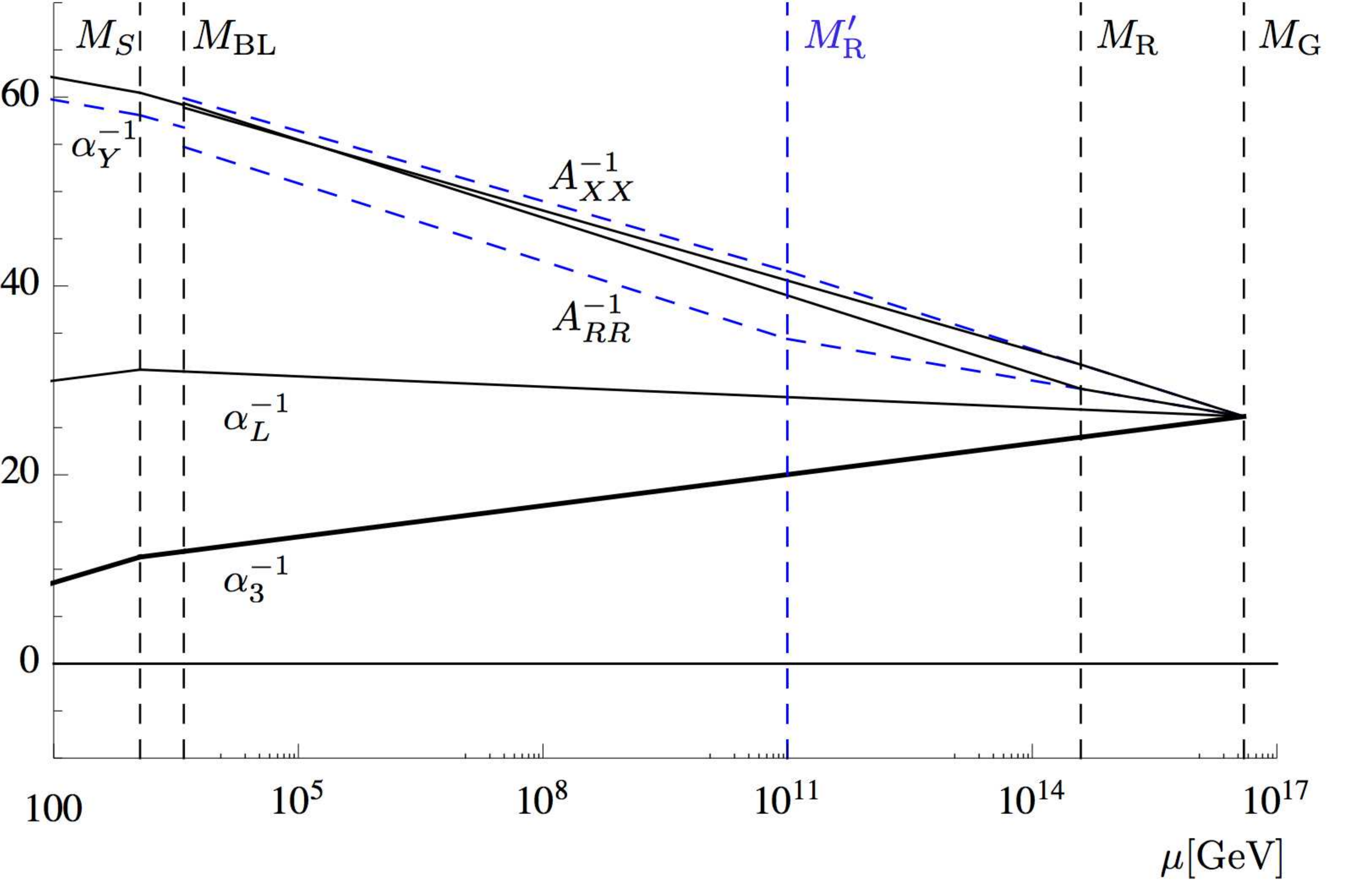}
\caption{The same like in FIG.~\ref{MRVrunningOK} but without the kinetic mixing effects taken into account. With the GUT-scale boundary condition and $M_{\rm R}$ as above, the low-energy value of $\alpha^{-1}_{Y}$, namely, $\alpha^{-1}_{Y}(M_{Z})=62.51$ (black solid lines) differs from the one obtained in the full-fledged calculation by as much as 4 percent. Alternatively, if one attempts to obtain the right value of  $\alpha^{-1}_{Y}(M_{Z})$ by adjusting the $SU(2)_{R}$-breaking scale, the new $M_{R}'$ scale must be  shifted with respect to the correct $M_{R}$ by as much as 4 orders of magnitude (in blue dashed lines). \label{MRVrunningWRONG}}
\end{figure}

However, this simple argument works only if the $U(1)$-mixing effects are 
properly taken into account. Remarkably, if they are simply
neglected, $\Delta\gamma$ receives only diagonal entries and 
$\alpha^{-1}_{Y}(M_{Z})$ becomes a function of $M_{\rm BL}$. 
Moreover, 
stretching the $M_{\rm BL}$-$M_{\rm R}$ range to maximum, the erroneous 
shift inflicted on $\alpha^{-1}_{Y}(M_{Z})$ can become as large as 4 
per-cent as can be seen by comparing figures \ref{MRVrunningOK} and
\ref{MRVrunningWRONG}. Alternatively, in order to retain the desired value 
of $\alpha^{-1}_{Y}(M_{Z})$, one would have to re-adjust $M_{\rm R}$ by several orders of magnitude, c.f., FIG.~\ref{MRVrunningWRONG}. This, however, could have a large impact on, e.g., the MSSM soft spectrum \cite{ValentinaMartinMichal}, and, in more general constructions, also on $M_{\rm G}$ and $\alpha_{G}$, with ramifications for $d=6$ proton decay etc. 

Finally, let us  note that the ``rotated-basis'' method discussed in brief in Sect.~\ref{sect:general} is only partially successful because the $g_{R}$ and $g_{B-L}$ gauge couplings do not coincide at the $M_{\rm R}$. Indeed, the value of  $\alpha^{-1}_{Y}(M_{Z})$ obtained this way, namely, $\alpha^{-1}_{Y}(M_{Z})=60.93$, is closer to the correct value than that received with no mixing at all, but still some 2\% off the correct value. 

\paragraph{Gaugino masses:} 
In order to fully appreciate the method advocated in this work, we should 
look at the interplay between the gauge and the soft sector. 
For example, at one loop-level, a simple illustration is by equation
(\ref{generalizedgauginogaugecorrelation}) which ties the gauge couplings 
$G$ together with the gaugino soft masses $M$. Consequently, the bino
mass  obeys at the scale $M_{\rm S}$
\be\label{binomass}
M_{Y}(M_{\rm S})=\frac{\alpha_{Y}(M_{\rm S})}{\alpha_{G}} p_{Y}^{T}m_{1/2}p_{Y}\,.
\ee 
where $m_{1/2}$ is the GUT-scale gaugino soft mass matrix.~From 
equation (\ref{binomass}) we see that the ratio 
$M_{Y}(M_{\rm S})/\alpha_{Y}(M_{\rm S})$ depends on whether one includes
the mixing effects or not as already noticed in ref.\ \cite{Kribs:1998rb}.
Note that with non-universal initial conditions, i.e.\ $m_{1/2}$ not
being proportional to the unit matrix, the $p_{Y}^{T}m_{1/2}p_{Y}$ term 
mixes up all entries of $m_{1/2}$. Moreover, in
 the special case that the abelian gauge couplings unify, even the one-loop gaugino sector evolution can be fully accounted for by the ``rotated-basis'' trick. 
\subsubsection{Two-loop effects}

At two-loop level our method becomes already important
in cases with gauge coupling unification at a certain scale.
We illustrate this by taking as an example the model presented
in ref.\ \cite{FileviezPerez:2010ek} where an intermediate 
$SU(3)_{c}\otimes SU(2)_{L} \otimes U(1)_Y \otimes U(1)_{B-L}$ gauge 
symmetry is assumed to originate from a grand-unified 
framework. 
We assume two cases: (i) full gauge coupling unification at
$2\times 10^{16}$GeV and (ii)  a small
difference of 5\% between the two $U(1)$ couplings caused by possible 
GUT-scale threshold effects. In the gaugino sector we assume
universal boundary conditions in both cases, but the effect gets
even stronger if one considers in addition threshold effects in the
gaugino sector as well.

The results are given in table~\ref{tab:comparison_running}. Remarkably,
besides the expected equivalence of the ``rotated-basis'' method and the 
 full-fledged calculation at the one-loop level, 
the relevant effective hypercharge gauge coupling turns
out to be identical to the one obtained even at two loop-level
if exact gauge couping unification is assumed. 
The reason is, that all additional states not present in the MSSM
are charged only with respect to  $U(1)_{B-L}$ but are neutral under
the MSSM gauge group. In the gaugino sector the first deviations
show up already in this case which however are only at the per-mile
level. In case that one includes also threshold corrections at the
GUT-scale the effects
are at the percent level leading to shifts in the masses
 potentially measurable
already at the LHC.

Last but not least we remark, 
that the effects would be even larger if the $U(1)_Y$ would
result from the breaking of $U(1)_R \otimes U(1)_{B-L}$ as discussed
in the previous example.

\begin{table*}[htb]
\begin{tabular}{|c|ccc|cc|cc|}
\hline
\hline
 & \multicolumn{3}{|c|}{One-loop results} & \multicolumn{4}{|c|}{Two-loop
results} \\
\hline
\parbox[0pt][3.5em][c]{0cm}{} & \parbox{2cm}{No kinetic mixing} &\parbox{2cm}{Rotated basis method} &
\parbox{2cm}{Complete RGEs} &  \parbox{2cm}{No kinetic mixing} &
\parbox{2cm}{Complete RGEs}&  \parbox{2cm}{No kinetic mixing} &
\parbox{2cm}{Complete RGEs} \\
\hline
\(g_{YY}\)       &  0.4511  & 0.4700 & 0.4700  &  0.4487  & 0.4677&  0.4487  & 0.4686\\
\(g_{BL BL}\)    &  0.4083  & 0.4243 & 0.4243  &  0.4070  & 0.4231&  0.4131  & 0.4298\\
\(g_{BL Y},g_{Y BL}\) &  0.  & -0.0723 & -0.0723 &  0.0   &-0.0725&  0.0   &-0.0725\\
\hline
$g_{Y}$  &  0.4511  & 0.4511 & 0.4511  &  0.4487  & 0.4487&  0.4487  & 0.4500\\
\hline
\(M_{Y Y}~[{\rm GeV}]\)   & 196.34   & 218.13 & 218.13  & 185.82   &
207.96& 185.80 & 208.71  \\
\(M_{BL BL}~[{\rm GeV}]\)     & 160.83   & 178.67 & 178.67 & 154.88  &
173.19& 144.26  & 161.97\\
\(M_{BL Y},M_{Y BL}~[{\rm GeV}]\)   & 0.0   & - 62.39 & - 62.39 & 0.0
&-63.10 & 0.0 &-62.15\\
\hline
$M_{Y}~[{\rm GeV}]$& 196.34   & 196.34    & 196.34     & 185.82   &185.96&
185.80   & 187.04  \\
\hline
& \multicolumn{5}{|c|}{Exact unification}  & \multicolumn{2}{|c|}{$g^{\rm
GUT}_{B L} = 1.05\, g^{\rm GUT}_Y$} \\ 
\hline
\hline
\end{tabular}
\caption{Low energy values of the entries of the gauge coupling and gaugino mass matrices ($g_{AB}$, $M_{AB}$) and the properly fitted MSSM parameters ($g_{Y}$, $M_{Y}$), c.f., Eqs.~(\ref{matching1diag}) and (\ref{matching2diag}). We
have fixed the GUT scale at $2 \times 10^{16}$ with $g_G = 0.72$ and imposed an
mSUGRA boundary condition taking $m_{1/2} = 500$~GeV. All gaugino mass
parameters are in GeV. 
At the one-loop level, we compare the case with no kinetic mixing effects
included, the ``rotated basis''  and the full-fledged calculation. 
At the two-loop level, we include the case where $g_Y$ and $g_{BL}$ 
are split  at the GUT scale due to threshold corrections.}
\label{tab:comparison_running}
\end{table*}

\section{Conclusions and outlook}
In this work, we have discussed the structure of the renormalization group 
equations in softly-broken supersymmetric models with more than a single 
abelian gauge group. 
Indeed, with multiple $U(1)$ gauge factors at play, the  effects of kinetic 
mixing among the abelian gauge fields must be taken into account in order 
to keep the theory renormalizable.

Though, formally, the evolution equations available in the literature do 
not exhibit any obvious pathologies if such subtleties are not taken into 
account, the calculations based on these formulas are in general
 incomplete and, 
thus, the results are internally inconsistent. This is even more
pronounced in the context of SUSY models because it affects also the 
evolution of the soft SUSY parameters, in particular the evolution
of the gaugino mass parameters.

Remarkably enough, the issue of the $U(1)$ mixing in the softly-broken
supersymmetric gauge theories has never been addressed in full generality, even
at one loop. The main aim of the current study was to fill this gap and provide
a fully self-consistent method for dealing with the renormalization group
evolution of the gauge couplings and the soft SUSY-breaking parameters up to the
two-loop level. 

To this end, we have studied in detail the existing two-loop 
renormalization group equations valid for the case of at most a single 
abelian gauge factor at play given in   ref.\
\cite{Martin:1993zk,Yamada:1994id} and 
extended these to account for the most general case of a gauge group with 
any number of  $U(1)$ factors.   

In particular, we have argued that all the $U(1)$ mixing effects can be 
consistently included if the  gauge couplings and the soft SUSY-breaking 
gaugino masses associated to the individual abelian gauge-group factors are 
generalized to matrices and these are then substituted into the formulae in 
\cite{Martin:1993zk,Yamada:1994id} in a specific manner. This, however,
is a highly non-
trivial enterprise, mainly  due to the non-commutativity of the 
relevant matrix-like structures, and a number of ambiguities had to be 
resolved. In this respect, the residual reparametrization invariance of the 
covariant derivative associated to the redefinition of the abelian gauge 
fields turned out to be a very useful tool, yet in many cases one had to 
resort to a detailed analysis of the relevant Feynman diagrams. 

The general method has been illustrated for two cases:
(i) at one-loop level where due to an breaking of the original
group two $U(1)$ factors emerge with different gauge couplings
and (ii) at the two-loop-level in a model where gauge coupling 
unifications occurs but where threshold corrections are taken into account.
In both case we obtain effects in the percent range and we remark,
that none of the previously proposed partial treatments can account for 
the full effects.

Last but not least, let us stress again that our results are 
completely generic and, as such, they do not require any specific 
assumptions about the charges of the chiral multiplets in the theory and/or 
the boundary conditions applied to the relevant gauge couplings. This makes 
the framework very suitable for implementation into computer
algebraic codes  calculating
two-loop renormalization group equations in softly-broken supersymmetric 
gauge theories such as {\tt SARAH}
\cite{Staub:2008uz,Staub:2009bi,Staub:2010jh} and {\tt Susyno}
\cite{Fonseca:2011sy}.

\section*{Acknowledgments}
We thank M.~Hirsch for fruitful discussion and F.~Braam and J.~Reuter for
interesting insights about the
different approaches to the kinetic mixing problem. In addition, we
thank B.~O`Leary for reading through the manuscript.

The work of M. M. was supported by the Marie Curie Intra European Fellowship
within the 7th European Community Framework Programme FP7-PEOPLE-2009-IEF,
contract number PIEF-GA-2009-253119,  by the EU Network grant UNILHC
PITN-GA-2009-237920, by the Spanish MICINN grants FPA2008-00319/FPA and
MULTIDARK CAD2009-00064 (Con-solider-Ingenio 2010 Programme) and by the
Generalitat Valenciana grant Prometeo/2009/091. W.P.\ and F.S.\ have been
supported by the
German Ministry of Education and Research (BMBF) under contract no.\ 05H09WWE. R.F. acknowledges the support provided by the Funda\c c\~ao para a Ci\^encia e a Tecnologia under the grant SFRH/BD/47795/2008.

\appendix
\section{Renormalization of QED $\otimes$ QED}\label{app:renormalizationscheme}
In this appendix, we comment in more detail on renormalization of abelian gauge
theories, focusing on the simplest non-trivial case exhibiting the effects of
kinetic mixing, namely the ``QED-squared'' scenario featuring two independent
abelian gauge groups $U(1)\otimes U(1)$.

Let us start with the basic bare Lagrangian of QED$^{2}$ including an explicit kinetic-mixing term 
\begin{eqnarray}
\nonumber {\cal
L}_{B} &= &
\overline{\psi}_{i,B}(i\slash\!\!\!\partial-m_{i,B})\psi_{i,B}-\overline{ \psi }
_{i,B}Q^T_i
G_{B}\slash\!\!\!\!A_{B}\psi_{i,B} \\
\label{bare:QED2} && \hspace{0.5cm} -\frac{1}{4}F_{B\mu\nu}\xi_{B}F_{B}^{
\mu\nu }\,.
\end{eqnarray}
Here $\psi_i$ are the relevant matter fields (whose number must
be in general equal to or greater than the number of the abelian gauge factors
otherwise there is no way to distinguish among all stipulated $U(1)$ factors),
$A$ stands for a 2-component vector (in the group space) comprising the gauge
fields associated to different $U(1)$ factors, $G$ stands for a (so far formal,
i.e, diagonal) $2\times 2$ containing the relevant pair of gauge couplings and
$\xi$ is a symmetric 
and real $2\times 2$ matrix parametrizing the gauge-kinetic form.  
\subsection{Scheme A: a non-canonical gauge propagator and diagonal gauge couplings}
Leaving $\xi_{B}$ in the game, one defines the renormalized and the counterterm Lagrangians as
\bea\label{ren:QED2A}
{\cal
L}\!&\!=
\!&\!\overline{\psi}_i(i\slash\!\!\!\partial-m_i)\psi_i-\overline{\psi}_i
Q^T_i G\slash\!\!\!\!A\psi_i-\tfrac{1}{4}F_{\mu\nu}\xi F^{\mu\nu}\,,
\\
\nn \delta{\cal L}\!&\!=\!&\! i\overline{\psi}_i\delta
Z_{\psi_i}\slash\!\!\!\partial\psi_i-\overline{\psi}_i\delta
Z_{m_i}m_i\psi_i-\overline{\psi}_i Q^T_i \delta Z_{G}G\slash\!\!\!\!A\psi_i \\
&& \hspace{0.5cm} -\tfrac{1}{4}
F_{\mu\nu}\delta \xi F^{\mu\nu}\!\!,
\eea
where
\bea
Q^T_i \delta Z_{G}G  &=& Z_{\psi_i} Q^T_i G_{B} Z_{A}^{1/2}- Q^T_i
G\,,\\
\delta \xi &=&  Z_{A}^{1/2}\xi_{B} Z_{A}^{1/2}-\xi \,, 
\label{deltaxi}
\eea
and $\delta Z_{m}$ and $\delta Z_{\psi_i}$ are 
unimportant for our considerations. These counterterms are fixed by 
the renormalization conditions so that they render the renormalized Green's 
functions of the theory UV-finite. For a diagonal $Z_{A}$ and for any fixed  
$\xi_{B}$, the off-diagonal entries
 in $\delta \xi$ cannot be matched by the 
right-hand side of Eq.~(\ref{deltaxi}) unless $\xi$ is a dynamical 
quantity. Remarkably, in this scheme the gauge coupling can be retained in 
a diagonal form throughout the RG evolution. This is because the relation 
between the bare and the renormalized couplings $G_{B}=GZ_{A}^{-1/2}$ can 
be brought into the form (trading $Z_{A}$ for $\xi$ and $\delta \xi$)
\be
G_{B}\xi_{B}^{-1}G_{B}^{T}=GZ_{A}^{-1/2}\xi_{B}^{-1} Z_{A}^{-1/2}G^{T}=G(\xi+\delta\xi)^{-1}G^{T}
\ee
(which holds to all orders in perturbation theory)
from where it is clear that any non-diagonal entry of the RHS of the 
evolution equation for $G$ can be absorbed into $\xi$. So, in this scheme, 
$\xi$ is a dynamical quantity while $G$ can be kept diagonal. 
\subsection{Scheme B: a canonical gauge propagator and non-diagonal gauge couplings}
If, instead, $\xi$ is absorbed by a suitable gauge-field 
redefinition $A\to \xi^{1/2}A\equiv \tilde A$  into  
matrix for the coupling constants, one is left with
\bea\label{ren:QED2B}
{\cal L}\!&\!=\!&\!
\overline{\psi}_i
(i\slash\!\!\!\partial-m_i)\psi_i-\overline{\psi}_i Q^T_i \tilde
G\slash\!\!\!\!\tilde A\psi_i-\tfrac{1}{4}\tilde F_{\mu\nu}\tilde F^{\mu\nu}\,,
\\
\nn \delta{\cal L}\!&\!=\!&\!  i\overline{\psi}_i \delta
Z_{\psi_i}\slash\!\!\!\partial\psi_i-\overline{\psi}_i\delta
Z_{m_i}m_i\psi_i-\overline{\psi}_iQ^T_i\delta Z_{\tilde G}\tilde
G\slash\!\!\!\!\tilde A\psi_i \\
&& \hspace{0.5cm} -\tfrac{1}{4} \tilde F_{\mu\nu}\delta Z_{\tilde
A} \tilde F^{\mu\nu}\!\!,\nn
\eea
where
\bea
Q^T_i \delta Z_{\tilde G}\tilde G  &=&  Z_{\psi_i} Q^T_i \tilde G_{B}
Z_{\tilde A}^{1/2}- Q^T_i \tilde G\,,\\
\delta Z_{\tilde A}&=&  (Z_{\tilde A}^{1/2})^{T}Z_{\tilde A}^{1/2}-1 \,, 
\label{deltaZAtilde}
\eea
with
$
Z_{\tilde A}^{1/2}= \xi^{-1/2}Z_{A}^{1/2}\xi_{B}^{1/2}
$
and, as before, 
\be
\tilde G_{B}=\tilde G Z_{\tilde A}^{-1/2}\,.
\ee 
It is again clear that the non-diagonality inflicted on $Z_{\tilde A}$ by the renormalization conditions renders the RHS of the gauge-coupling evolution equation non-diagonal. However, in this scheme, $\xi$ has been swallowed by the gauge-field renormalization counterterm and, as such, does not need to be treated as an extra dynamical quantity. In other words, the whole effect is accounted for by the off-diagonal form of the generalized gauge coupling $\tilde G$.

\section{Recapitulation of the two-loop RGEs for simple groups and their products with at most  one $U(1)$}\label{sect-rges}
\subsection{Case A: Simple gauge
group\label{sect-rges-simple}}
For completeness we display
 here the RGEs in the case of a simply gauge group
based on \cite{Martin:1993zk,Yamada:1994id,Jack:1994kd}. 
For a  general $N=1$ supersymmetric gauge theory with superpotential  
\begin{equation}
 W (\Phi) = L_i \Phi_i + \frac{1}{2}{\mu}^{ij}\Phi_i\Phi_j + \frac{1}{6}Y^{ijk}
\Phi_i\Phi_j\Phi_k \thickspace ,
\end{equation}
the  soft SUSY-breaking scalar terms are given by
\begin{align}
\nonumber V_{\rm soft} = & \left(S^i \phi_i +   \frac{1}{2}b^{ij}\phi_i\phi_j
+ \frac{1}{6}h^{ijk}\phi_i\phi_j\phi_k +\hbox{c.c.}\right) \\
& +(m^2)^i{}_j\phi_i\phi_j^* + \frac{1}{2} M \lambda_a \lambda_a
\thickspace.
\end{align}
Here we will follow \cite{Martin:1993zk} and assume that repeated indices are summed over.
Note also that lowered indices imply conjugation (e.g., $Y_{ijk}\equiv{Y^{ijk}}^*$).
In the notation defined in sect.~\ref{sect-results}, the anomalous dimensions
of the chiral superfields are given by 
\begin{align}
\label{example1}
 & \gamma_i^{(1)j} =  \frac{1}{2} Y_{ipq} Y^{jpq} - 2 \delta_i^j g^2 C(i)
\thickspace ,\\
& \gamma_i^{(2)j}  =  g^2
Y_{ipq} Y^{jpq} [2C(p)- C(i)]  -\frac{1}{2} Y_{imn} Y^{npq} Y_{pqr} Y^{mrj}
\nonumber \\
 & \hspace{0.05\linewidth}  + 2 \delta_i^j g^4 [ C(i) S(R)+ 2 C(i)^2 - 3 C(G)
C(i)] \thickspace ,
\end{align}
and the \(\beta\)-functions for the gauge couplings are given by
\begin{align}
 \beta_g^{(1)}  =  &  g^3 \left[S(R) - 3 C(G) \right] \thickspace , \\
\nonumber \label{g5SRCR} \beta_g^{(2)}  =  &  g^5 \left\{ - 6[C(G)]^2 + 2 C(G) S(R) + 4 S(R)
C(R) \right\} \\
&     - g^3 Y^{ijk} Y_{ijk}C(k)/d(G) \thickspace .
\end{align}
The corresponding RGEs are defined as
\begin{equation}
\frac{d}{dt} g  =   \frac{1}{16\pi^2} \beta_g^{(1)} +  \frac{1}{(16\pi^2)^2}
\beta_g^{(2)} \thickspace.
\end{equation}
Here, we used \(t=\ln Q\), where \(Q\) is the renormalization scale. The
$\beta$-functions for the superpotential parameters can be obtained by using
superfield technique. The obtained expressions are 
\begin{eqnarray}\nonumber
 \beta_Y^{ijk} &= & Y^{ijp} \left [
 \frac{1}{16\pi^2}\gamma_p^{(1)k} +
 \frac{1}{(16\pi^2)^2}  \gamma_p^{(2)k} \right ] \\
&& + (k \leftrightarrow i) + (k\leftrightarrow j) \thickspace , \\
\nonumber \beta_{\mu}^{ij} &= & \mu^{ip} \left [
 \frac{1}{16\pi^2}\gamma_p^{(1)j} +
 \frac{1}{(16\pi^2)^2} \gamma_p^{(2)j} \right ]
+ (j \leftrightarrow i) \thickspace , \\
\\
\beta_L^i & = & L^{p} \left [
 \frac{1}{16\pi^2}\gamma_p^{(1)i} +
 \frac{1}{(16\pi^2)^2}  \gamma_p^{(2)i} \right ] \thickspace .
\end{eqnarray}
The expressions for trilinear, soft-breaking terms are
\begin{align}
\frac{d}{dt} h^{ijk}  =  & \frac{1}{16\pi^2} \left [\beta^{(1)}_h\right ]^{ijk}
+  \frac{1}{(16\pi^2)^2} \left [\beta^{(2)}_h\right ]^{ijk} \thickspace ,
\end{align}
with
\begin{align}
& \left [\beta^{(1)}_h\right ]^{ijk}   =
  \frac{1}{2} h^{ijl} Y_{lmn} Y^{mnk}
+ Y^{ijl} Y_{lmn} h^{mnk} \\
\nonumber & \hspace{0.05\linewidth}- 2 \left (h^{ijk} - 2 M Y^{ijk}  \right )
g^2  C(k)  + (k \leftrightarrow i) + (k \leftrightarrow j) \thickspace , \\
& \left [\beta^{(2)}_h\right ]^{ijk}   =
-\frac{1}{2} h^{ijl} Y_{lmn} Y^{npq} Y_{pqr} Y^{mrk} \nonumber \\
& \hspace{0.05\linewidth}- Y^{ijl} Y_{lmn} Y^{npq} Y_{pqr} h^{mrk} - Y^{ijl}
Y_{lmn} h^{npq} Y_{pqr} Y^{mrk} \nonumber \\
& \hspace{0.05\linewidth}+ \Big ( h^{ijl} Y_{lpq} Y^{pqk} +  2 Y^{ijl} Y_{lpq}
h^{pqk} \nonumber \\
& \hspace{0.2\linewidth}- 2 M Y^{ijl} Y_{lpq} Y^{pqk} \Big ) g^2\left[ 2 C(p) -
C(k) \right ] \nonumber \\
& \hspace{0.05\linewidth}+ \left (2h^{ijk} - 8 M Y^{ijk} \right )
g^4 \Big [  C(k)S(R)+ 2 C(k)^2  \nonumber \\
& \hspace{0.2\linewidth}- 3 C(G)C(k) \Big ] + (k \leftrightarrow i)
+ (k \leftrightarrow j)  \thickspace .
\end{align}
For the bilinear soft-breaking parameters, the expressions read
\begin{align}
\label{example2}
\frac{d}{dt} b^{ij}  =  &
 \frac{1}{16\pi^2}\left [\beta^{(1)}_b \right ]^{ij} +
 \frac{1}{(16\pi^2)^2} \left [\beta^{(2)}_b \right ]^{ij} \thickspace , 
\end{align}
with
\begin{align}
\label{example4}
& \left [\beta^{(1)}_b  \right ]^{ij}   = 
\frac{1}{2} b^{il} Y_{lmn} Y^{mnj} +\frac{1}{2}Y^{ijl} Y_{lmn} b^{mn}
\nonumber \\
& \hspace{0.05\linewidth} + \mu^{il} Y_{lmn} h^{mnj}   - 2 \left (b^{ij} - 2 M
\mu^{ij} \right )g^2 C(i) 
 + (i \leftrightarrow j) \thickspace , \\
& \left [\beta^{(2)}_b \right ]^{ij}   = 
 -\frac{1}{2} b^{il} Y_{lmn} Y^{pqn} Y_{pqr} Y^{mrj} \nonumber \\
& \hspace{0.05\linewidth}-\frac{1}{2} Y^{ijl} Y_{lmn} \mu^{mr} Y_{pqr} h^{pqn}
- \mu^{il} Y_{lmn} h^{npq} Y_{pqr}  Y^{mrj} \nonumber \\
& \hspace{0.05\linewidth} - \mu^{il} Y_{lmn} Y^{npq}  Y_{pqr} h^{mrj} 
 - \frac{1}{2} Y^{ijl} Y_{lmn} b^{mr} Y_{pqr} Y^{pqn} \nonumber \\
& \hspace{0.05\linewidth} + 2 Y^{ijl} Y_{lpq} \left ( b^{pq} - \mu^{pq} M \right
) g^2 C(p)
 + \Big ( b^{il} Y_{lpq} Y^{pqj} \nonumber \\
& \hspace{0.05\linewidth} + 2 \mu^{il} Y_{lpq}
h^{pqj}
- 2 \mu^{il} Y_{lpq} Y^{pqj} M \Big )
g^2 \left[ 2 C(p) - C(i) \right ]  \nonumber \\
& \hspace{0.05\linewidth} +  \left ( 2 b^{ij} - 8 \mu^{ij} M\right )
g^4 \Big [ C(i)S(R)+ 2 C(i)^2 \nonumber \\
& \hspace{0.2\linewidth} - 3 C(G)C(i)   \Big ] + (i \leftrightarrow j) 
\thickspace , 
\end{align}
Finally, the RGEs for the linear soft-breaking parameters are
\begin{align}
\frac{d}{dt} S^{i}  =  &
 \frac{1}{16\pi^2}\left [\beta^{(1)}_S \right ]^{i} +
 \frac{1}{(16\pi^2)^2} \left [\beta^{(2)}_S \right ]^{i} \thickspace ,
\end{align}
with
\begin{align}
& \left [\beta^{(1)}_S \right]^i  =
\frac{1}{2}Y^{iln}Y_{pln}S^{p}
+L^{p}Y_{pln}h^{iln}
\nonumber \\
& \hspace{0.05\linewidth}+\mu^{ik}Y_{kln}b^{ln}+2Y^{ikp}(m^2)_{p}^l\mu_{kl}
+h^{ikl}b_{kl} \thickspace ,\\
& \left [\beta^{(2)}_S \right]^i  = 
2g^2C(l)Y^{ikl}Y_{pkl}S^{p} -\frac{1}{2}Y^{ikq}Y_{qst}Y^{lst}Y_{pkl}S^{p}
\nonumber \\
& \hspace{0.05\linewidth}-4g^2C(l)(Y^{ikl}M-h^{ikl}) Y_{pkl}L^{p} 
\nonumber \\
& \hspace{0.05\linewidth} -\big[Y^{ikq}Y_{qst}h^{lst}Y_{pkl}
+h^{ikq}Y_{qst}Y^{lst}Y_{pkl}\big]L^{p}
\nonumber \\
& \hspace{0.05\linewidth} -4g^2C(l)Y_{jnl} (\mu^{nl}M-b^{nl})\mu^{ij} 
-\big[Y_{jnq}h^{qst}Y_{lst}\mu^{nl} \nonumber \\
&  \hspace{0.05\linewidth} +Y_{jnq}Y^{qst}Y_{lst}b^{nl}\big]\mu^{ij}
+4g^2C(l)(2Y^{ikl}\mu_{kl}|M|^2 \nonumber \\
&  \hspace{0.05\linewidth}  -Y^{ikl}b_{kl}M
-h^{ikl}\mu_{kl} M^* +h^{ikl}b_{kl}\nonumber \\
&  \hspace{0.05\linewidth} +Y^{ipl}(m^2)_{p}^k\mu_{kl}
 +Y^{ikp}(m^2)_{p}^l\mu_{kl}) 
\nonumber \\
 & \hspace{0.05\linewidth}-\Big[Y^{ikq}Y_{qst}h^{lst}b_{kl}
+h^{ikq}Y_{qst}Y^{lst}b_{kl}
  \nonumber \\
 & \hspace{0.05\linewidth}+h^{ikq}h_{qst}Y^{lst}\mu_{kl} 
+Y^{ipq}(m^2)_{p}^kY_{qst}Y^{lst}\mu_{kl}\nonumber \\
 & \hspace{0.05\linewidth}
+Y^{ikq}Y_{qst}Y^{pst}(m^2)_{p}^l\mu_{kl}
+Y^{ikp}(m^2)_{p}^qY_{qst}Y^{lst}\mu_{kl}\nonumber \\
 & \hspace{0.05\linewidth}+2Y^{ikq}Y_{qsp}(m^2)_t^{p}Y^{lst}\mu_{
kl} +Y^{ikq}h_{qst}h^{lst}\mu_{kl} \Big] \thickspace .
\end{align}
With these results, the list of the \(\beta\)-functions for all couplings is
complete. Now, we turn to  the RGEs for the gaugino masses, squared masses of
scalars and vacuum expectation values. The result for the gaugino masses is 
\begin{align}
 \frac{d}{dt} M = & \frac{1}{16 \pi^2} \beta_M^{(1)}
+ \frac{1}{(16 \pi^2)^2 } \beta_M^{(2)} \thickspace ,
\end{align}
with
\begin{align}
\label{example5}
\beta_M^{(1)} = & g^2 \left[ 2 S(R) - 6 C(G) \right] M \thickspace ,
\\
\nonumber \beta_M^{(2)}
= & g^4\left\{ -24[C(G)]^2 + 8 C(G) S(R) + 16  S(R)C(R)\right\} M\\
 & + 2 g^2 \left [h^{ijk}  - M Y^{ijk}\right] Y_{ijk}
C(k)/d(G) \thickspace .
\end{align}
The one- and two-loop RGEs for the scalar mass parameters read
\begin{align}
\frac{d}{dt} \left(m^2\right)_i^j  =  &
 \frac{1}{16\pi^2} \left [\beta^{(1)}_{m^2} \right ]_i^j +
 \frac{1}{(16\pi^2)^2} \left [\beta^{(2)}_{m^2} \right ]_i^j \thickspace , 
\end{align}
with
{\allowdisplaybreaks
\begin{align}
& \left [\beta^{(1)}_{m^2} \right ]_i^j  =  
 \frac{1}{2} Y_{ipq} Y^{pqn} {(m^2)}_n^j
+ \frac{1}{2} Y^{jpq} Y_{pqn} {(m^2)}_i^n \nonumber \\
 & \hspace{0.05\linewidth}+ 2 Y_{ipq} Y^{jpr} {(m^2)}_r^q
 + h_{ipq} h^{jpq}  - 8 \delta_i^j M M^\dagger g^2 C(i) \nonumber \\
 & \hspace{0.05\linewidth}+
 2 g^2 {\bf t}^{Aj}_i {\rm Tr} [ {\bf t}^A m^2 ] \thickspace , \\
& \left [\beta^{(2)}_{m^2} \right ]_i^j  =  
 -\frac{1}{2} {(m^2)}_i^l Y_{lmn} Y^{mrj} Y_{pqr} Y^{pqn}
 \nonumber \\
 & \hspace{0.05\linewidth}-\frac{1}{2} {(m^2)}^j_l Y^{lmn} Y_{mri} Y^{pqr}
Y_{pqn}  \nonumber \\
 & \hspace{0.05\linewidth}
- h_{ilm} Y^{jln} Y_{npq} h^{mpq} - Y_{ilm} Y^{jnm} {(m^2)}_n^r Y_{rpq} Y^{lpq}
\nonumber  \\
& \hspace{0.05\linewidth} - Y_{ilm} Y^{jnr} {(m^2)}_n^l Y_{pqr} Y^{pqm}
- Y_{ilm} Y^{jln} h_{npq} h^{mpq}  \nonumber \\
 & \hspace{0.05\linewidth}- 2 Y_{ilm} Y^{jln}  Y_{npq} Y^{mpr} {(m^2)}_r^q -
h_{ilm} h^{jln} Y_{npq} Y^{mpq} \nonumber\\
&  \hspace{0.05\linewidth}- Y_{ilm} Y^{jnm}
{(m^2)}_r^l Y_{npq} Y^{rpq} - Y_{ilm} h^{jln}
h_{npq} Y^{mpq}
\nonumber \\
 & \hspace{0.05\linewidth} + \biggl [{(m^2)}_i^l Y_{lpq} Y^{jpq}
+ Y_{ipq} Y^{lpq} {(m^2)}_l^j + 4 Y_{ipq} Y^{jpl} {(m^2)}_l^q
\nonumber \\
 & \hspace{0.05\linewidth} +  2 h_{ipq} h^{jpq}
 - 2 h_{ipq} Y^{jpq} M -2 Y_{ipq} h^{jpq} M^\dagger
\nonumber \\
 & \hspace{0.05\linewidth} + 4Y_{ipq} Y^{jpq} M M^\dagger
\biggr ]
g^2 \left [C(p) + C(q)- C(i) \right ]  \nonumber \\
 & \hspace{0.05\linewidth}-2 g^2 {\bf t}^{Aj}_i ({\bf t}^A m^2)_r^l
Y_{lpq} Y^{rpq}
+ 8 g^4 {\bf t}^{Aj}_i {\rm Tr} [ {\bf t}^A C(r) m^2 ]  \nonumber \\
& \hspace{0.05\linewidth}+ \delta_i^j g^4 M M^\dagger \Big [
24C(i) S(R) + 48 C(i)^2 - 72 C(G) C(i) \Big ] \nonumber \\
 & \hspace{0.05\linewidth}
 + 8 \delta_i^j g^4 C(i) ( {\rm Tr} [S(r) m^2] - C(G) M M^\dagger ) \thickspace
.
\label{example3}
\end{align}}
The RGEs for a VEV \(v^i\) are proportional to the anomalous dimension of the
chiral superfield whose scalar component receives the VEV
\begin{equation}
 \frac{d}{dt }v^i =  v^{p} \left [
 \frac{1}{16\pi^2}\gamma_p^{(1)i} +
 \frac{1}{(16\pi^2)^2}  \gamma_p^{(2)i} \right ]
\end{equation}

\subsection{Product groups with at most one $U(1)$\label{sect-rges-products}} 
To generalize the formulas above to the case of a
direct product of gauge groups, the following substitution rules are needed
\cite{Martin:1993zk}. Note, we give these replacements here only for
completeness and they are not sufficient in the case of several $U(1)$ gauge
groups, see sec.~\ref{sect-results} for the necessary extensions. \\ 
For the \(\beta\) functions of gauge couplings and gauginos the rules are
\allowdisplaybreaks
\begin{align}
g^3 C(G)  &\rightarrow g_a^3 C(G_a) \thickspace ,\\
g^3 S(R)  &\rightarrow  g_a^3 S_a(R)  \thickspace ,\\
g^5 C(G)^2  &\rightarrow  g_a^5 C(G_a)^2 \thickspace ,\\
g^5 C(G) S(R)  &\rightarrow  g_a^5 C(G_a) S_a(R) \thickspace ,\\
g^5 S(R) C(R)  &\rightarrow  \sum_b g_a^3 g_b^2 S_a(R) C_b(R) \thickspace ,\label{g5SRCRsubst}\\
16 g^4 S(R) C(R) M  &\rightarrow  8 \sum_b g_a^2 g_b^2 S_a(R) C_b(R) (M_a + M_b)
\thickspace ,\\
g^3 C(k)/d(G)  &\rightarrow  g_a^3 C_a(k)/d(G_a) \thickspace .
\end{align}
For all the other \(\beta\) functions, we need
\begin{align}
\label{example1product}
 g^2 C(r) &\rightarrow \sum_a g_a^2 C_a(r) \thickspace , \\
 g^4 C(r) S(R) &\rightarrow \sum_a g_a^4 C_a(r) S_a(R) \thickspace , \\
 g^4 C(r) C(G) &\rightarrow \sum_a g_a^4 C_a(r) C(G_a)\thickspace , \\
 g^4 C(r)^2 &\rightarrow \sum_a \sum_b g_a^2 g_b^2 C_a(r) C_b(r)\thickspace , \\
\nonumber 48 g^4 M M^\dagger C(i)^2 &\rightarrow \sum_a \sum_b g_a^2 g_b^2
C_a(i) C_b(i) \times \\
& \hspace{-0.1\linewidth} \times (32 M_a M_a^\dagger + 8 M_a M_b^\dagger + 8 M_b
M_a^\dagger) \thickspace ,\\
g^2 {\bf t}_i^{A j} \mbox{Tr}({\bf t}^A m^2) &\rightarrow \sum_a g_a^2 ({\bf t}_a^A)^j_i
\mbox{Tr}({\bf t}_a^A m^2)\thickspace , \\
g^2 {\bf t}_i^{A j} ({\bf t}^A m^2)^l_r Y_{lpq} Y^{rpq} &\rightarrow \sum_a g_a^2
({\bf t}_a^A)^j_i ({\bf t}_a^A m^2)^l_r Y_{lpq} Y^{rpq}\thickspace , \\
g^4 {\bf t}_i^{A j} \mbox{Tr}[{\bf t}^A C(r) m^2] &\rightarrow \sum_a \sum_b g_a^2 g_b^2
\nonumber ({\bf t}_a^A)^j_i \times \\
& \hspace{0.1\linewidth} \times \mbox{Tr}[{\bf t}_a^A C_b(r) m^2]\thickspace , \\
g^4 C(i) \mbox{Tr}[S(r) m^2] &\rightarrow \sum_a g_a^4 C_a(i) \mbox{Tr}[S_a(r)
m^2] \thickspace .
\end{align}
\section{Obtaining the substitution rules}\label{sect-appendix-results}
In this Appendix, we illustrate in more detail the methods used throughout the derivation of the substitution rules given in section \ref{sect-results}.

\subsection{The role of the $V_{i}$ vectors and the $M$ matrix
}

As mentioned in the text, the $U(1)$ gauge coupling matrix $G$ and
the charge vectors $Q_{i}$ of the chiral superfields $\Phi_{i}$
appear always through the combination $V_{i}=G^{T}Q_{i}$. The only
exception are the RGEs of $G$, where there should be a leading free
$G$. For example, the $\psi^{i\dagger}\psi^{i}A_{\mu}^{a}$ vertex
is proportional to $V_{i}^{a}$ (component $a$ of the vector $V_{i}$),
\be
\parbox{4cm}{
\includegraphics[scale=0.8]{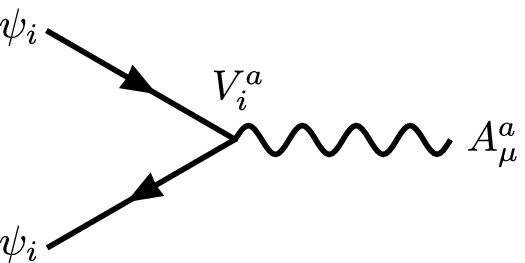}}
\ee
Similarly the vertices $\phi^{*i}\phi^{i}A_{\mu}^{a}$, $\phi^{*i}\phi^{i}A_{\mu}^{a}A_{\nu}^{b}$,
$\phi^{i*}\psi^{i}\lambda^{a}$ and the Yukawa independent part of
$\phi^{i*}\phi^{i}\phi^{j*}\phi^{j}$ are proportional to $V_{i}^{a}$,
$V_{i}^{a}V_{i}^{b}$, $V_{i}^{a}$ and $V_{i}^{T}V_{j}$ respectively.
In addition, there is to consider the $U(1)$ gaugino mass matrix $M$,
\be
\parbox{3cm}{\includegraphics[scale=0.8]{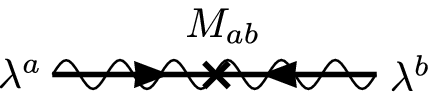}}
\ee
where $M_{ab}$ is the $a,b$ component of $M$.

\subsection{RGEs with no $U(1)$ indices }

Only diagrams underlying the RGEs for $G$ and $M$ contain external
$U(1)$ gauge bosons/gauginos. As such, in all other equations, while
vectors $V_{i}$ and the matrix $M$ may be present, they must be
in combinations that are scalars with no free $U(1)$ indices.

Consider $Mg^{2}C\left(i\right)$ appearing in the one-loop RGE of
the bilinear scalar soft terms $b^{ij}$, which is to be replaced
by $\sum_{A}M_{A}g_{A}^{2}C_{A}\left(i\right)+V_{i}^{T}MV_{i}$. The
simple groups contribution does not interest us though, so we shall
neglect it. We can see that $V_{i}^{T}MV_{i}$ is the only structure
that can generalize the expression $Mg^{2}C\left(i\right)=Mg^{2}y_{i}^{2}$
for one $U(1)$ group only. Observe also the contraction of the $U(1)$
indices in the expression - it comes from the possibility of having
any of the $U(1)$ gauginos in the internal lines of the contributing
diagram,
\be
\parbox{5cm}{\includegraphics[scale=0.8]{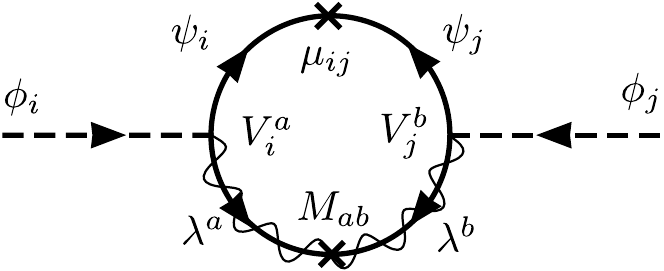}}
\ee
The amplitude is proportional to $\sum_{a,b}V_{i}^{a}M_{ab}V_{j}^{b}\mu^{ij}=\mu^{ij}V_{i}^{T}MV_{j}$.
Note that, for any pair of values $i,j$ the gauge symmetry forces
$\mu^{ij}=0$ unless $V_{i}+V_{j}=0$ which means that $\mu^{ij}V_{j}=-\mu^{ij}V_{i}$
so the amplitude of the diagram is indeed proportional to $V_{i}^{T}MV_{i}$.

This requirement that expressions with $V$'s and $M$'s must form
scalars is enough to derive the Eqs.~(\ref{rule34})-(\ref{rule17}),
(\ref{rule20})-(\ref{rule27}) and 
(\ref{rule30}) from the existing substitution
rules for gauge groups with multiple factors. We are left with the
terms $g^{4}C\left(r\right)S\left(R\right)$, $Mg^{4}C\left(r\right)S\left(R\right)$,
$g^{4}C\left(i\right)\textrm{Tr}\left[S\left(r\right)m^{2}\right]$
and $24g^{4}MM^{*}C\left(i\right)S\left(R\right)$. Note that  
one can
write $S\left(R\right)$ as $\textrm{Tr}\left[S\left(r\right)\right]$
in the notation of ref.\ 
\cite{Martin:1993zk}, so in all four cases there is
a sum over field components of chiral superfields. For diagrams with
up to two-loops and with no external gauginos nor gauge bosons, the
factors $S\left(R\right)$ and $\textrm{Tr}\left[S\left(r\right)m^{2}\right]$
can only come from the following sub-diagrams%
\footnote{It is conceivable that they could come also from diagrams with one
$\phi^{*}\phi^{*}\phi\phi$ vertex, but we may choose an appropriate
gauge, the Landau gauge, where these are 0 because an external scalar
line always couples to a gauge bosons at a three point vertex.%
}:
\be
\parbox{7.5cm}{\includegraphics[scale=0.6]{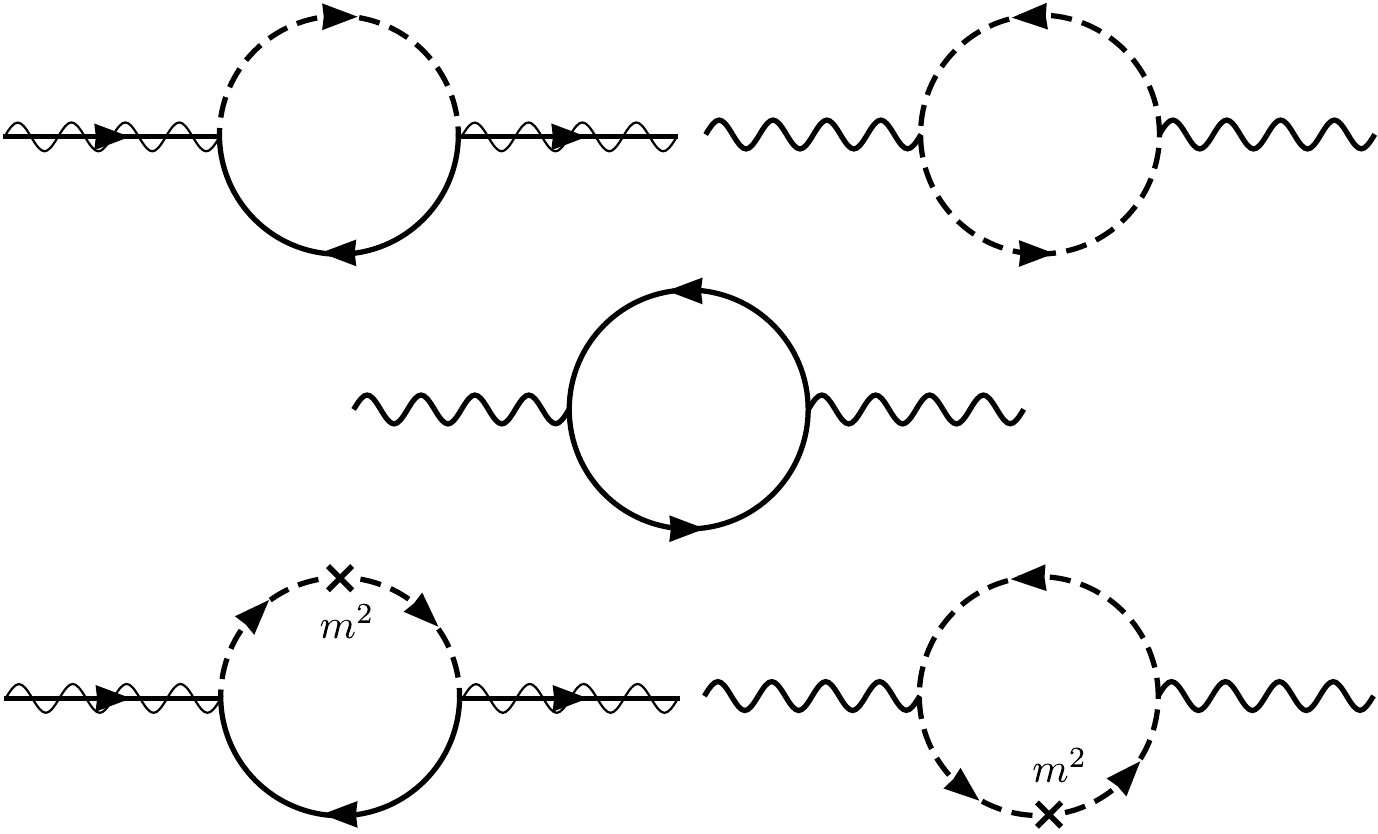}}
\ee
Take for example $g^{4}C\left(r\right)S\left(R\right)\sim\sum_{p}g^{4}y_{r}^{2}y_{p}^{2}$.
The reason why one cannot immediately
generalize this expression to include
$U(1)$ mixing effects is because in theory it could take the form
$\sum_{p}\left(V_{r}^{T}V_{p}\right)\left(V_{r}^{T}V_{p}\right)$
or $\sum_{p}\left(V_{r}^{T}V_{r}\right)\left(V_{p}^{T}V_{p}\right)$.
But looking at the above diagrams, such ambiguities go away because
 in all cases the $V$'s which are summed over (the $V_{p}$'s)
do not contract with each other. They contract with something else
at the other ends of the gauge boson/gaugino lines. As such, there
are no $V_{p}^{T}V_{p}$'s in these expressions and with this piece
of information, combined with
the known rules for a gauge group with multiple factors, 
Eqs.~(\ref{rule18}), (\ref{rule19}) and (\ref{rule28})
follow.  The substitution
rule given in Eq.~(\ref{rule29}) for
$24g^{4}MM^{*}C\left(r\right)S\left(R\right)$
appearing in the two-loop equation of the soft scalar masses is more
complicated since the placement of the $M$, $M^{\dagger}$ gaugino
mass matrices between these $V$'s is relevant. Nevertheless, from
the following diagrams we can calculate that the $U(1)$'s contribution
to this term is $8\sum_{p}\left[\left(V_{i}^{T}MV_{p}\right)\left(V_{i}^{T}M^{\dagger}V_{p}\right)+\left(V_{i}^{T}MM^{\dagger}V_{p}\right)\left(V_{i}^{T}V_{p}\right)\right.$
$\left.+\left(V_{i}^{T}M^{\dagger}MV_{p}\right)\left(V_{i}^{T}V_{p}\right)\right]$:

\begin{align}
& \parbox{5cm}{\includegraphics[scale=0.8]{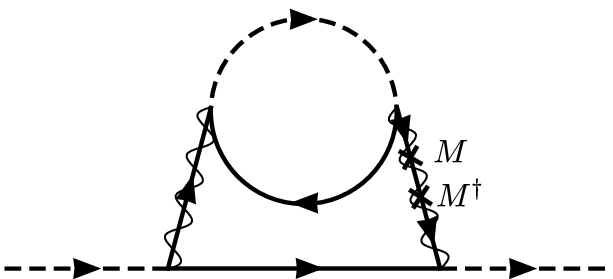}}\\
& \parbox{5cm}{\includegraphics[scale=0.8]{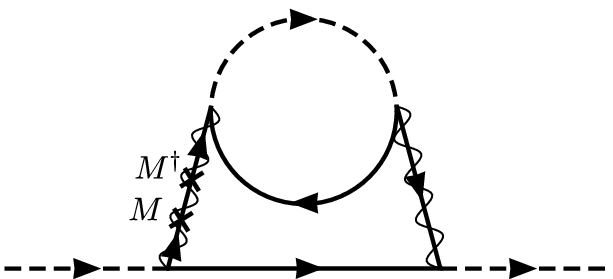}}\\
& \parbox{5cm}{\includegraphics[scale=0.8]{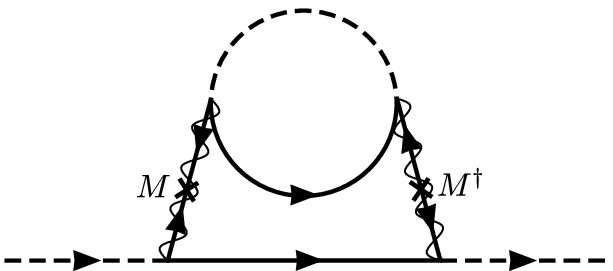}}
\end{align}

\subsection{RGEs with $U(1)$ indices}

The RGEs for $G$ and $M$ are the only ones with free $U(1)$ indices.
For the beta functions of the gaugino masses, we will be interested
in looking at diagrams with two incoming gauginos. As for the coupling
constant, due to the Ward identities, the contributing diagrams are
those with two external gauge bosons. From the amplitude of these
diagrams we still have to add a $G$ factor in order to obtain $\beta_{G}$.
Pictorially,

\be
\parbox{7cm}{\includegraphics[scale=0.7]{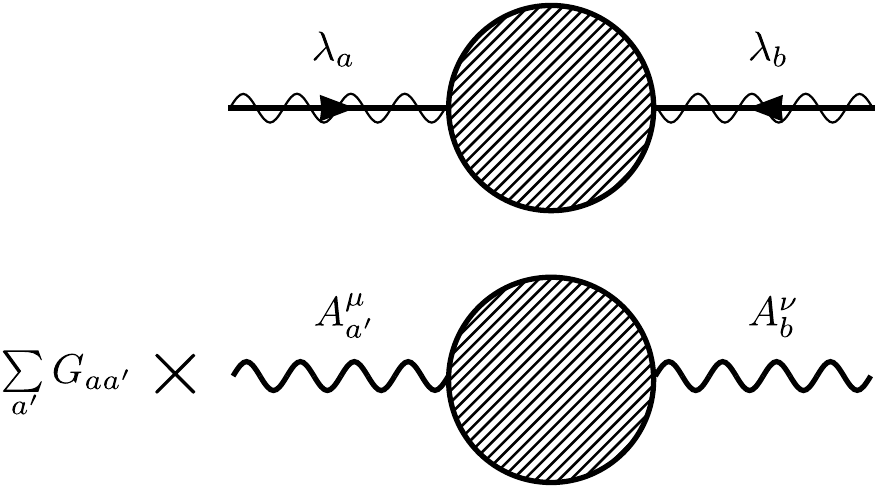}}
\ee
Note that all the terms in $\beta_{G}$ must be of the form
$GV_{i}^{T}\left(\cdots\right)V_{j}$ for some $i,j$ as mentioned
in the text. These qualitative considerations suffice for our purposes.

Keep also in mind that
\begin{itemize}
\item[(i)] the RGEs are invariant under the set of transformations $G\rightarrow O_{1}GO_{2}^{T}$,
$V_{i}\rightarrow O_{2}V_{i}$, $M\rightarrow O_{2}MO_{2}^{T}$ for
any orthogonal matrices $O_{1}$, $O_{2}$;
\item[(ii)] $M$ is a symmetric matrix and so must be $\frac{dM}{dt}$.
\end{itemize}
Taken together, these considerations allow us to deduce Eqs.~
(\ref{rule32})-(\ref{rule40}) (Eq.~(\ref{rule31}) is trivial).

We shall exemplify this for the case of
$16g^{4}S\left(R\right)C\left(R\right)M$
which for multiple factor groups is replaced by
$8\sum_{b}g_{a}^{2}g_{b}^{2}S_{a}\left(R\right)C_{b}\left(R\right)\left(M_{a}+M_{b}\right)$
in the RGEs of $M_{a}$. This is the same as $8\sum_{p,b}g_{a}^{2}g_{b}^{2}\frac{S_{a}\left(p\right)C_{b}\left(p\right)}{d_{a}\left(p\right)}\left(M_{a}+M_{b}\right)$.
Groups $a$ and $b$ are independent so the expressions $\sum_{b}g_{b}^{2}C_{b}\left(p\right)$,
$\sum_{b}M_{b}g_{b}^{2}C_{b}\left(p\right)$ are decoupled from $g_{a}^{2}\frac{S_{a}\left(p\right)}{d_{a}\left(p\right)}$,
$M_{a}g_{a}^{2}\frac{S_{a}\left(p\right)}{d_{a}\left(p\right)}$.
Inclusion of $U(1)$ mixing effects in the first pair of expressions
is easy because there are no free $U(1)$ indices:
$\sum_{b}g_{b}^{2}C_{b}\left(p\right)\rightarrow\sum_{B}g_{B}^{2}C_{B}
\left(p\right)+V_{p}^{T}V_{p}$
and $\sum_{b}M_{b}g_{b}^{2}C_{b}\left(p\right)\rightarrow\sum_{B}M_{B}g_{B}^{2}C_{B}\left(p\right)+V_{p}^{T}MV_{p}$.
If the group $a$ is an $U(1)$, then in case of a single
$U(1)$ this corresponds to 
$g_{a}^{2}\frac{S_{a}\left(p\right)}{d_{a}\left(p\right)}=g^{2}y_{p}^{2}$
which generalizes to
$g_{a}^{2}\frac{S_{a}\left(p\right)}{d_{a}\left(p\right)}\rightarrow
V_{p}V_{p}^{T}$.
Similarly, the only symmetric matrix expression which respects
the $O_{2}$ symmetry that can generalize $M_{a}g_{a}^{2}\frac{S_{a}\left(p\right)}{d_{a}\left(p\right)}$
is $\frac{1}{2} (MV_{p}V_{p}^{T} + V_{p}V_{p}^{T}M)$.

Assembling these pieces gives Eqs.~\ref{rule37} for
$16g^{4}S\left(R\right)C\left(R\right)M$.
The structure of the final expression is verifiable by looking at
the relevant diagrams:
\be
\parbox{7.5cm}{\includegraphics[scale=0.61]{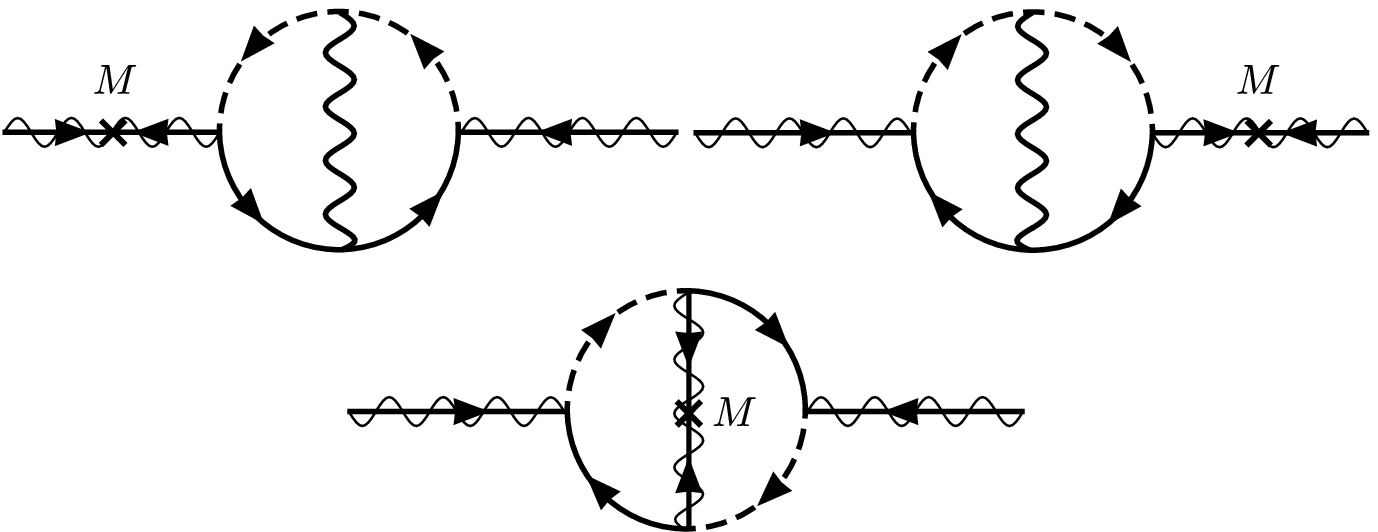}}
\ee

\section{Gauge-coupling and gaugino-mass matching}
\label{app:matching}
In this appendix we comment on yet another key ingredient of any practical application of the methods advocated in this work, namely, on the matching between the symmetric and asymmetric phases of spontaneously broken models with multiple $U(1)$ gauge groups, where the latter is described by an effective theory with a reduced dimensionality of the abelian sector.   
\subsection{Gauge couplings}
Whenever a charged field develops a VEV, the original $U(1)_{a}\otimes U(1)_{b}$ gauge symmetry gets broken down to a single $U(1)_{c}$ spanned over the unbroken combination of the original generators 
\be\label{ps}
Q^{c}_i=p_{a}Q^{a}_i+p_{b}Q^{b}_i\;\;\; \text{(no summation).}
\ee
A gauge coupling associated to this residual symmetry is then in general a function of the original gauge couplings $g_{a}$ and $g_{b}$,  the $p$-coefficients above, and, at higher loop order, also other quantities such as group Casimirs etc. 
 Let us recapitulate in brief how the matching between the two regimes 
determines $g_{c}$, focusing mainly on the ``tree-level matching'' as needed
when working with the one-loop RGEs. We comment on the changes
for  two-loop evolution at the end of this section.

It is convenient to address the issue in two steps. First, one can consider the
matching between two QED$^{2}$ scenarios which are connected just by a pair of
$O$-rotations as in Eqs.~(\ref{O1})-(\ref{O2}). In particular, any specific
choice of $O_{1}$ represents a transition from the original set of the $U(1)$
charges  $Q_i=(Q^{a}_i,Q^{b}_i)^{T}$ to a new set 
${Q'}_i=(Q^{c}_i,Q^{d}_i)^{T}$ where ${Q'}_i=O_{1}Q_i$. This, in turn,
transforms the relevant (matrix of) gauge couplings as $G\to G'=O_{1}G$ so that
the interaction part of the covariant derivative remains intact,
$Q_i^{T}GA={Q'}_i^{T}G' A$. Regardless of whether simultaneously an
$O_{2}$ rotation has been performed on the gauge fields, i.e., $A\to A'=O_{2}A$
(inducing $G\to G'=O_{1}GO_{2}^{T}$), one always has $G'G'^{T}=O_{1}G G
^{T}O_{1}^{T}$, or, equivalently, 
\be\label{matching}
(G'G'^{T})^{-1}=O_{1}(G G ^{T})^{-1}O_{1}^{T}
\ee
which, as we shall see below, is more useful in practice.
Hence,  for any specific choice of $O_{1}$, this equation yields the link between the specific non-linear combinations of $g_{aa}$, $g_{ab}$, $g_{ba}$ and $g_{bb}$ entries of $G$ and the  $g_{cc}$, $g_{cd}$, $g_{dc}$ and $g_{dd}$ entries of $G'$ and, as such, provides the desired matching condition. Note that, given the $p_{a}$ and $p_{b}$ coefficients in Eq.~(\ref{ps}), the first row of $O_{1}$ is fixed and its second row is determined from orthogonality up to a global sign.

However, $O_{2}$ plays an important role if one, e.g., needs to get a coupling associated to a specific gauge 
field, for instance the one associated to the ``residual'' $U(1)_{c}$ that survives the $U(1)_{a}\otimes U(1)_{b}$ 
breakdown. This gauge boson corresponds to the massless eigenstate of the gauge-boson mass matrix 
\be
M_{A}^{2}=G^{T}\langle H \rangle^{\dagger} Q^{*}_H Q_H^{T}\langle H \rangle G
\ee
where $\langle H \rangle$is the $U(1)_{a}\otimes U(1)_{b}$-breaking VEV. 
Since  the conserved charge $Q^{c}_H$ in Eq.~\ref{ps} annihilates this VEV, it is convenient to go into the
primed basis where ${Q'}_H^{T}\langle H \rangle=(0,V)$. Thus,
\begin{eqnarray}
\nn M_{A}^{2}&=&G^{T}O_{1}^{T}\langle H \rangle^{\dagger}Q'^{*}_H {Q'^{T}_H}\langle H
\rangle O_{1} G \\
&=&
G^{T}O_{1}^{T}\left(
\begin{array}{cc}
0 & 0 \\
0 & V^{2}
\end{array}
\right)O_{1} G
\end{eqnarray}
A convenient gauge-field transformation $A\to A'=O_{2}A$ would bring this mass matrix into a diagonal form 
\be\label{MAdiagonal}
M_{A'}^{2}=
G'^{T}\left(
\begin{array}{cc}
0 & 0 \\
0 & V^{2}
\end{array}
\right)G'
\ee
(using $G'=O_{1}GO_{2}^{T}$) if and only if $G'$ is upper-triangular, i.e.,
\be\label{Gprimetriangular}
G'=\left(
\begin{array}{cc}
g'_{cc} & g'_{cd} \\
0 & g'_{dd}
\end{array}
\right)
\ee
 The zero in the 12 entry  has a clear physical implication: 
for the couplings of the surviving (massless) $A'_{c}$ gauge boson
only  $g_{c}\equiv g'_{cc}$ is relevant in the effective theory whereas
one does need to care the $g'_{cd}$ and $g'_{dd}$ couplings of the heavy 
$A'_{d}$ which is integrated out except for the calculation of
higher-dimensional effective operators.
 
From here it is also easy to understand why Eq.~(\ref{matching}) is better suited for practical purposes than the one without inverse: Indeed, the 1-1 entry of $(G'G'^{T})^{-1}$ reveals $g'_{cc}$ in a very simple way, namely,
\be\label{matching1diag}
(G'G'^{T})^{-1}=\left(
\begin{array}{cc}
{g'^{-2}_{cc}} & -g'_{cd} {g'^{-1}_{dd}}{g'^{-2}_{cc}}\\
. & (g'^{2}_{cc}+g'^{2}_{cd}) {g'^{-2}_{dd}}{g'^{-2}_{cc}}
\end{array}
\right)
\ee
(the dotted term follows from symmetry) so that it is sufficient to look at the 1-1 entry of the RHS of Eq.~(\ref{matching}), whilst without the inverse 
\begin{equation}
G'G'^{T}=
\left(
\begin{array}{cc}
g'^{2}_{cc}+g'^{2}_{cd}& g'_{cd}g'_{dd} \\
. & g'^{2}_{dd}
\end{array}
\right)
\end{equation}
and in order to extract $g'_{cc}$ one has to solve a non-linear system involving all three independent entries of the RHS of Eq.~(\ref{matching}), namely, $O_{1}GG^{T}O_{1}^{T}$. 

At higher-loop orders, the situation becomes slightly more involved, 
especially if the $U(1)\otimes U(1)$ gauge structure is tensored with a 
semi-simple gauge factor $G_X$.  For example gauge-boson 
canonical normalization effects  have to be considered at the 
two-loop level \cite{Weinberg:1980wa,Hall:1980kf}, which yields 
for example extra 
group-Casimir factors associated to $G_X$ entering formulae like 
Eq.~(\ref{matching}), see, e.g., \cite{Bertolini:2009qj}. The specific 
shape of these terms is, however, renormalization scheme dependent.
 
\subsection{SUSY and the gaugino masses}
Concerning the gaugino masses, the situation is only slightly more involved here. Sticking to the simplest $U(1)_{a}\otimes U(1)_{b}\to U(1)_{c}$ case as before, the gaugino($\lambda$)-higgsino($\tilde H$) mass matrix reads schematically (in the $O_{2}$-rotated basis bringing the gauge boson mass matrix to a block-diagonal form, c.f., Eq.(\ref{MAdiagonal}), and with $O_{1}$ rotation imposed on charges)  
\be
M_{\lambda',\tilde H}=
\left(
\begin{array}{ccc}
M'&\sqrt{2} G'^{T}{Q'_H}\langle H\rangle^{*} & \sqrt{2} G'^{T}{Q'_H}\langle H^{c}\rangle^{*} \\
. & W_{HH} & W_{HH^{c}}\\
. & .& W_{H^{c}H^{c}}\\
\end{array}
\right)\,.
\ee
Here $M'=O_{2}MO_{2}^{T}$ is the gaugino soft mass matrix, $W$ is the superpotential; the subscripts of $W$ denote the derivatives of $W$ with respect to the superfield $H$ and its charge conjugate $H^{c}$. 
As before, due to the specific choice of the primed basis one has
 ${Q'^{T}_H}\langle H \rangle=(0,V)$ and ${Q'}_H^{T}\langle H^{c}
\rangle=(0,V^{*})$. Furthermore, the triangular shape of $G'$, see
Eq.~(\ref{Gprimetriangular}), ensures that the $\lambda_{c}$ gaugino
corresponding to the first row/column receives no mass contribution due to the
spontaneous symmetry breaking.

Thus, one concludes that the effective soft mass of the gaugino associated to the surviving gauge group $U(1)_{c}$ is given by the $c$-$c$ entry of the $M'$ mass matrix 
\be
M_{c}=M'_{cc}\,.
\ee
Note that, in principle, there is no need to calculate the $O_{2}$ matrix because one can trade it for the gauge couplings and the (known) $O_{1}$:
\be
O_{2}M O_{2}^{T}=G'^{T} O_{1}G^{-1T}MG^{-1}O_{1}^{T}G'
\ee
and thus
\be\label{gauginomatchingmatrix}
G'^{-1T}M'G'^{-1}=O_{1}G^{-1T}MG^{-1}O_{1}^{T}\,.
\ee
The main advantage of this formula is that, again, the LHS reveals $(M')_{cc}$  in a particularly simple manner, namely
\be\label{matching2diag}
G'^{-1T}M'G'^{-1}= \left(
\begin{array}{cc}
{M'_{cc}}/{g'^{2}_{cc}} & R_{12}\\
. & R_{22}
\end{array}
\right)\,,
\ee
with $R_{12}\equiv {-g'_{cd}M'_{cc}+g'_{cc}M'_{cd}
}{{g'_{dd}}/{g'^{2}_{cc}}}$ and $R_{22}\equiv {(g'^{2}_{cc}M'_{dd}-2g'_{cc}g'_{cd}M'_{cd}+g'^{2}_{cd}M'_{cc})}/{{g'^{2}_{dd}}{g'^{2}_{cc}}}$,
and, thus, in combination with (\ref{matching1diag}),  admits for a simple extraction of the surviving gaugino effective soft mass. Moreover, due to the one-loop RGE invariance of the $G^{-1T}MG^{-1}$ combination,  the RHS of Eq.~(\ref{gauginomatchingmatrix}) is directly connected to the initial condition.

\bibliographystyle{h-physrev5}
\bibliography{bib}
\end{document}